\documentclass[aps,pre,floatfix,twocolumn,showpacs]{revtex4-1}
\usepackage{color}
\usepackage{graphicx}
\usepackage{amsmath, amsthm, amssymb}

\newcommand{\be}{\begin{equation}}
\newcommand{\ee}{\end{equation}}
\newcommand{\bea}{\begin{eqnarray}}
\newcommand{\eea}{\end{eqnarray}}

\begin{document}
\title{Multiple Phase Transitions in Extended Hard Core Lattice Gas
Models in Two Dimensions}
\author{Trisha Nath}
\email{trishan@imsc.res.in}
\affiliation{The Institute of Mathematical Sciences, C.I.T. Campus,
Taramani, Chennai 600113, India}
\author{R. Rajesh}
\email{rrajesh@imsc.res.in}
\affiliation{The Institute of Mathematical Sciences, C.I.T. Campus,
Taramani, Chennai 600113, India}
\date{\today}

\begin{abstract}
We study the $k$-NN hard core lattice gas model in which the first $k$ 
next nearest neighbor sites of a particle are excluded from occupation 
by other particles on a two dimensional square lattice. This model is 
the lattice version of the hard disc system with increasing $k$ 
corresponding to decreasing lattice spacing. While the hard disc system is 
known to undergo a two step freezing process with increasing density, 
the lattice model has been known to show only one transition. Here, based
on Monte Carlo simulations and high density expansions of the free energy and
density, we 
argue that for $k=4,10,11,14...$, the lattice model undergoes multiple 
transitions with increasing density. Using Monte Carlo simulations, we 
confirm the same for $k=4,\ldots, 11$. This, in turn, resolves an 
existing puzzle as to why the 4-NN model has a continuous transition 
against the expectation of a first order transition. 
\end{abstract}


\maketitle

\section{\label{I}Introduction}

Lattice gas models of particles that interact only through excluded 
volume interactions are among the simplest systems that undergo phase 
transitions -- all order-disorder transitions being entropy 
driven~\cite{runnels1}. They are closely related to the freezing 
transition~\cite{alderdisc,aldersphere}, directed and undirected lattice 
animals~\cite{deepak2,deepak1,imbrie}, and the Yang-Lee edge 
singularity~\cite{parisi}. The different particle shapes that have been 
studied include squares~\cite{bellemans_nigam1nn,pears,ramola}, 
hexagons~\cite{baxter}, dimers with nearest neighbor exclusion~\cite{dickman},
triangles~\cite{verberkmoes}, 
tetrominoes~\cite{barnes}, rods~\cite{deepak,kundu2}, 
and rectangles~\cite{kundu3}. Despite sustained interest, exact solutions 
exist only for the hard hexagon and related 
models~\cite{baxter,bouttier}. Thus, it is worthwhile to study such 
models using numerical methods and qualitative arguments to understand 
the dependence of the phase diagram on the shape of the particle.

In this paper, we focus on the $k$-NN hard core lattice gas model 
(HCLG), where the first $k$ next nearest neighbors of a particle may not 
be occupied by another particle, on the two dimensional square lattice 
(see Sec.~\ref{II} for a precise definition for the model).  Introduced 
by Domb and Burley in the 1950s~\cite{domb,burley1,burley2}, the $k$-NN HCLG model has found 
applications in diverse areas of research. Examples include adsorption on 
surfaces~\cite{taylor1985,patrykiejew,Mitchell,koper1,koper2,bak,zhang,bartelt}, 
limiting cases of spin 
models~\cite{amar,landau,landau1985}, frustrated antiferromagnets at 
high magnetic fields~\cite{derzhko,zhi2}, glass transitions on 
square~\cite{eisenberg2000,eisenberg2005} and Bethe 
lattices~\cite{weigt},
study of two dimensional Rydberg gases~\cite{ji2011}, 
and in combinatorial problems~\cite{baxter1999} such as unfriendly 
theater sitting problem~\cite{georgiou}, random independent set problem 
on graphs~\cite{scott_sokal}, loss networks~\cite{suhov}, 
q-coloring graphs~\cite{sokal2000personal} and reconstruction 
problems~\cite{bhatnagar}.

The $k$-NN HCLG is also the lattice version of the hard-sphere problem in 
the continuum, where larger k corresponds to smaller 
lattice spacing. The hard sphere system undergoes an entropy driven 
transition to a solid phase at high densities. In two dimensions, the generally accepted 
KTHNY scenario predicts two continuous transitions: first from a liquid 
phase to a hexatic phase with quasi long range orientational order and 
second from the hexatic phase to a solid phase with quasi long range 
positional order and long range orientational 
order~\cite{kosterlitz,nelson,yong}. The order of these transitions, however, 
continue to be debated (see Refs.~\cite{krauth2011,wierschem2011} and 
references within for a recent discussion).

On the other hand, the $k$-NN model for $k \leq 5$ is known to exhibit 
only one transition with increasing density~\cite{fernandes}.  The 1-NN 
model has been shown to undergo a continuous phase transition of the 
Ising universality class from a low density disordered phase to a high 
density sublattice ordered phase by series 
expansion~\cite{gaunt_fisher,bellemans_nigam1nn,baxter_enting_tsang,baram}, 
transfer matrix method~\cite{bellemans_nigam1nn,runnels,combs,ree2,nisbet,guo,pears,chan1,jensen}, Monte 
Carlo simulations~\cite{landau,meirovitch,hu_mak,fernandes,liu}, density 
functional theory~\cite{lafuente}, and renormalization group 
methods~\cite{racz,hu_chen}.  For 2-NN (the $2 \times 2$ hard square model), 
in contrast to 1-NN, the sublattice ordered phase has a sliding 
instability that makes the high density phase columnar. In this phase, 
translational order is present along either rows or columns but not 
along both~\cite{bellemans_nigam1nn,bellemans_nigam3nn,ree,landau,nisbet1,lafuente2003phase,ramola,schmidt,amar,slotte,kinzel}. Monte 
Carlo simulations show that the disorder--columnar transition in the 
2-NN model is continuous and belongs to the two color Ashkin Teller 
universality class -- the order parameter for the 2-NN model having a 
four fold symmetry. Recent work estimating the critical exponents may be 
found in Refs.~\cite{fernandes,feng,zhi2,ramola2}. For higher values of 
$k$, the number of symmetric high density ordered states are $10$ 
(3-NN), $8$ (4-NN) and $6$ (5-NN). By analogy with the $q$-state Potts 
model, it is expected that the transitions to an ordered state in these 
models are first order. Indeed, all evidence shows that the 3-NN model undergoes a 
first order phase transition from a low density disordered phase
into a sublattice phase with 
increasing density~\cite{bellemans_nigam3nn,orban1966,orban1982,bellemans_nigam1nn,nisbet2,heilmann,eisenberg2005,fiore}. 
The 5-NN model is equivalent to the $3 \times 3$ hard square problem and 
the high density phase is columnar as in the 
2-NN model~\cite{fernandes,nisbet1}. The transition in the 5-NN model
has been numerically shown to be first order~\cite{fernandes},
though very early studies claimed absence of a phase transition~\cite{nisbet1}. However, 
for the 4-NN model, 
Monte Carlo simulations, Mayer cluster integral analysis, and
transfer matrix methods
show a surprise~\cite{fernandes,rotman,nisbet2}. Rather than a 
first order transition, the 4-NN model was shown to undergo a continuous 
transition.
The critical exponents obtained from Monte Carlo simulations are 
indistinguishable
from those of the two dimensional Ising model~\cite{fernandes}. However, the analysis based on
cluster integrals excludes the possibility of the transition belonging to the 
Ising universality class~\cite{rotman}. Early transfer matrix studies
suggested weak first order or continuous transition~\cite{nisbet2}.
Thus, the nature of the transition in the 4-NN model
has remained a puzzle.

Not much is known for $k \geq 6$. It becomes increasingly difficult to 
equilibrate systems with large $k$ in Monte Carlo simulations that use 
only local evaporation, deposition and diffusion moves. At high densities, 
when the excluded volume of a particle is large, the system gets stuck 
in long lived metastable states. Thus, reliable data can be obtained
only for low densities or small excluded volumes. Mean field approximations 
predict single continuous transitions for $k=1,2$ and single first order 
transitions for $k>2$~\cite{domany}.

Does the $k$-NN model show multiple transitions like its continuum 
counterpart? What is the rationalization for 4-NN undergoing a 
continuous transition rather than a first order transition? In this 
paper, we address these questions by adapting and implementing an 
efficient algorithm with cluster moves~\cite{kundu} that has proved 
very useful in studying high density regimes of systems with large 
excluded volume like long hard rods~\cite{kundu2}, hard 
rectangles~\cite{kundu3} and hard squares~\cite{ramola2}. Using this
algorithm, we are able to numerically study systems up to $k=11$, a
significant increase from the earlier studies up to  
$k=5$~\cite{fernandes,domany,nisbet1,nisbet2,rotman}. The algorithm 
is explained in detail, along with the definition of the model, in 
Sec.~\ref{II}. For the 4-NN model, we show that the system undergoes two 
continuous transitions with increasing density and that the high density 
phase is columnar.  This, in effect, resolves the question of why the 
4-NN model showed a continuous transition by arguing that the eight fold 
symmetry of the model is broken in two steps. The exponents describing 
the two transitions are numerically determined. The first transition is 
consistent with the Ising universality class while the second transition 
has exponents that belong to the two color Ashkin Teller model. The 
numerical study of the 4-NN model is presented in Sec.~\ref{III}. In 
Sec.~\ref{IV}, we calculate the first four terms in the high density 
expansion for the free 
energy and the densities of particles in the different sublattices  of the 4-NN
model. From the form of the expansion, it is seen that the columnar
order has a sliding instability in only some sublattices. This
observation is  used to heuristically argue why the system shows two
entropy driven  transitions. In Sec.~\ref{V}, we generalize the arguments 
for the 4-NN model, 
based on the high density expansion, to larger $k$ and conjecture a 
criteria for multiple transitions to be observed with increasing 
density.  In particular, we argue that for a fixed $k$, 
if the high density phase is 
columnar but the sliding instability is not along all sublattices, then 
the system should undergo multiple transitions with increasing density. 
Applying this criteria to larger 
$k$, we argue that the HCLG with $k=10,11,14,\ldots$ should undergo 
multiple transitions while $k=6,7,8,9$ should have a single first order 
transition. In Sec.~\ref{VI}, we present results from Monte Carlo 
simulations for $6\leq k \leq 11$. It is shown that for $k=6,7,8,9$, 
there is a single first order transition. For $k=10$, we show that there 
are two transitions-- one continuous and the other first order. The
exponents describing the continuous transition are shown to be consistent 
with those of the two dimensional Ising model. For 
$k=11$, we show that there are at least two transitions with increasing 
density. Section~\ref{VII} contains a discussion of the results and some
possible extensions of the problem.

\section{\label{II}Model and Monte Carlo Algorithm}

\begin{figure}
\includegraphics[width=0.9 \columnwidth]{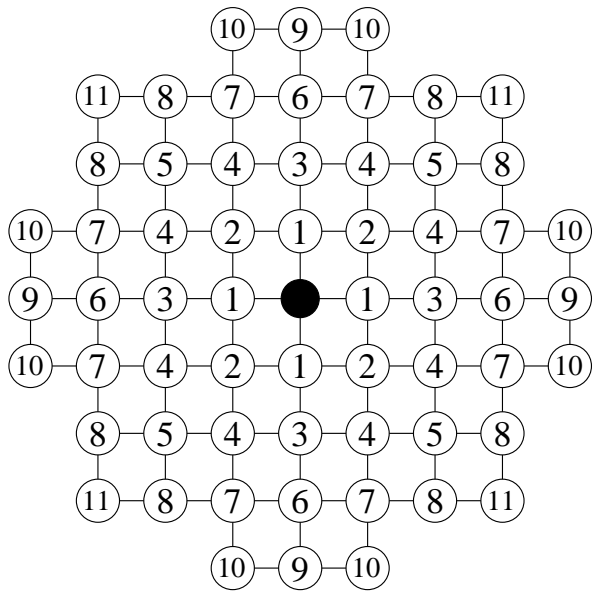}
\caption{The lattices sites that are excluded by a particle (black 
filled circle). The labels denote the sites that are the $k^{th}$ next 
nearest neighbors.  For $k$-NN HCLG, all sites with labels less than or 
equal to $k$ are excluded.
}
\label{fig:knn}
\end{figure}

Consider a square lattice of size $L \times L$ with periodic boundary 
conditions. A lattice site may be occupied by utmost one particle. The 
first $k$ next nearest neighbors of a particle are excluded from being 
occupied by another particle. This corresponds to all lattice sites 
within a distance $R$ where $R^2$ is a positive integer. In 
Fig.~\ref{fig:knn}, the sites excluded by a particle are shown for 
$k=1,2,\ldots, 11$. For a given $k$, all sites with labels less than or 
equal to $k$ are excluded. An activity $z=\exp(\mu)$ is associated with 
each particle, where $\mu$ is the chemical potential.

We study this system using a grand canonical Monte Carlo algorithm. The 
algorithm is an adaptation of an efficient algorithm with cluster moves 
well suited to study hard core problems~\cite{kundu,kundu2}.

\begin{figure}
\includegraphics[width=\columnwidth]{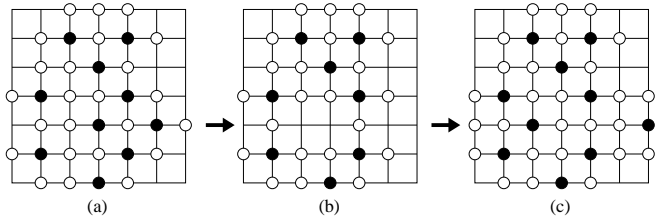}
\caption{The Monte Carlo algorithm illustrated for 1-NN. (a) A typical 
valid configuration. The lattice consists of particles (filled circles), 
excluded sites (empty circles) and empty sites where a particle may be 
added to obtain a new valid configuration. (b) A row is chosen at random 
(denoted by arrow) and all particles on that row are removed. The row is 
now divided into intervals of contiguous empty sites by the excluded 
sites. (c) The row is reoccupied with a new valid configuration with the 
correct equilibrium weight.
}
\label{fig:mc}
\end{figure}

We describe the implementation of the algorithm for the 1-NN model and 
then outline the modifications required to implement it for larger $k$. 
Consider a valid configuration of the 1-NN model [see Fig.~\ref{fig:mc} 
(a)]. A row or column is chosen at random (say a row) and all particles 
on that row are removed. 
The aim is to reoccupy the row with a new configuration with the correct 
equilibrium weight. 
After evaporation, the row is divided by the excluded sites 
into intervals of contiguous empty sites [see Fig.~\ref{fig:mc} (b)]. In the 1-NN 
model, along a row, a particle excludes the nearest neighbor from being 
occupied by a particle. Thus, the particle configuration in an interval 
is independent of its neighboring intervals, and the re-occupation of 
the row reduces to the occupation of empty intervals.

The calculation of probability of a configuration may be determined from 
the exact solution of the one dimensional problem on a lattice of length 
$l$ with open and periodic boundary conditions. Let $\Omega_o(z,\ell)$ 
[$\Omega_p(z,\ell)$] denote the partition function of the problem with 
nearest neighbor exclusion on a lattice with open [periodic] boundary 
conditions. They obey recursion relations
\begin{subequations}\label{omega}
\begin{eqnarray}
\Omega_o(z,\ell)&=&1+\ell z (1-\delta_{\ell,0}),~ \ell=0,\ldots, d,\\
\Omega_o(z,\ell)&=&z \Omega(z,\ell-d-1)+\Omega(z,\ell-1),~ \ell>d,\\
\Omega_p(z,L)&=&d z\Omega_o(z,L-2d-1)+\Omega_o(z,L-d),
\end{eqnarray} 
\end{subequations}
where for the 1-NN model, $d=1$. Given an empty interval of length $\ell 
<L$, the probability that the left most site is occupied equals $z 
\Omega_{z,\ell-d}/\Omega_\ell$. If $\ell=L$, then the probability that one 
of the first $d$ sites is occupied equals $d z \Omega_{z,L-d-2}/ 
\Omega_p(z,L)$. These probabilities are calculated for all $\ell$ and 
stored as input for the Monte Carlo simulations. A Monte Carlo step 
corresponds to $2 L$ such evaporation--deposition moves. It is 
straightforward to show that the algorithm is ergodic and obeys detailed 
balance.

\begin{figure}
\includegraphics[width=\columnwidth]{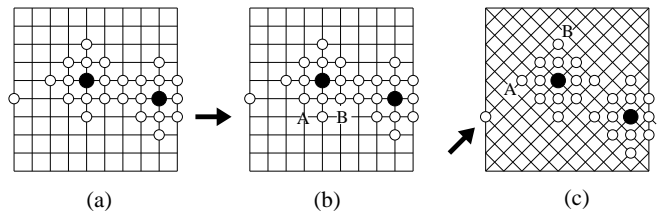}
\caption{An example of the algorithm for 3-NN. (a) A valid configuration of 
2 particles (filled circles) and excluded sites (empty circles). (b) 
Adding particles in a row (denoted by arrow) by deposition. If 
a particle is added at A, then its exclusion range includes B. Thus, the 
occupation of the empty intervals are not independent. (c) Adding
particles in a diagonal oriented in the $\pi/4$ direction 
(denoted by arrow). Now the occupation of the empty intervals are 
independent of each other.}
\label{fig:3nn_move}
\end{figure}

The algorithm is easily generalized to higher values of $k$. However, one 
cannot always choose rows and columns for the evaporation--deposition 
moves because the occupation of the empty intervals may no longer be 
independent of one another. This is most easily seen for the 3-NN model, 
where along a row, a particle excludes the nearest and next nearest 
neighbors. However, in a row, only one site is excluded by a particle 
that is two rows away. An example is shown in Fig.~\ref{fig:3nn_move}. A 
valid configuration is shown in Fig.~\ref{fig:3nn_move}(a). If a 
deposition is attempted in the row denoted by an arrow [see 
Fig.~\ref{fig:3nn_move}(b)], then the occupation of site A excludes site 
B which belongs to a different empty interval. This makes the occupation 
of different empty intervals interdependent. However, if one attempts 
evaporation and deposition along diagonals oriented in the $\pi/4$ 
direction [Fig.~\ref{fig:3nn_move}(c)], then the occupation of empty 
intervals become independent of each other. In Table~\ref{table:param}, 
we tabulate the orientations of the diagonals for the 
evaporation--deposition moves that we have used for $k$ up to $k=11$. 
For each of these choices, the values of $d$ that should be used in 
Eq.~(\ref{omega}) are also tabulated in 
Table~\ref{table:param}. 
\begin{table}
\caption{\label{table:param} For each $k$, the orientation of the 
diagonals along which particles are evaporated and deposited, and the corresponding 
value of $d$ to be used in Eq.~(\ref{omega}) are tabulated. Evaporation and deposition 
along diagonals separated by $\Delta$ or more are independent.}
\begin{ruledtabular}
\begin{tabular}{cllc}
k & orientation of diagonal  & d & $\Delta$\\ \hline
   1 &  $0$, $\pi/2$  & 1 & 2\\
   2 &  $0$, $\pi/2$  & 1 & 2\\ 
   3 &  $\pi/4$, $3\pi/4$ & 1 & 3\\ 
   4 &  $0$, $\pi/2$ & 2 & 3\\ 
   5 &  $0$, $\pi/2$  & 2 & 3\\ 
   6 &  $\tan^{-1}(\pm2)$,  $\tan^{-1}(\pm1/2)$ & 1 & 7\\ 
   7 &  $\pi/4$, $3\pi/4$ & 2 & 5\\ 
     &   $0$, $\pi/2$  & 3 & 4\\ 
   8 &  $\pi/4$,  $3\pi/4$ & 2 & 6\\ 
     &   $0$, $\pi/2$  & 3 & 4 \\ 
   9 &  $\pi/4$, $3\pi/4$ &  2 & 6\\ 
  10 &  $\pi/4$,  $3\pi/4$ &  2 & 6\\ 
  11 & $\tan^{-1}(\pm2)$,  $\tan^{-1}(\pm1/2)$ & 1 & 10
\end{tabular}
\end{ruledtabular}
\end{table}

We implement a parallel version of the above algorithm. The evaporation 
and deposition of particles in two rows in the 1-NN model  (diagonals in general)
that have at least $\Delta -1$ rows between them
($\Delta=2$ for the 1-NN model) 
are independent of each other. The value of $\Delta$ for different $k$ are
given in Table~\ref{table:param}. Hence, we update simultaneously 
every $\Delta$th row. Once all rows are updated in this manner, 
the columns are updated. The parallelization and efficiency of the 
algorithm allows us to simulate large system sizes and high densities.

We check for equilibration by initializing the simulations with two 
different initial configurations, corresponding to two different phases, 
and making sure that the final equilibrium state is independent of the 
initial condition. One configuration is a fully packed state and the 
other is a random configuration where particles are deposited at random.

In a typical run for a fixed value of $\mu$, after equilibration, the 
different thermodynamic quantities are averaged over $10^8$ Monte Carlo 
steps that are divided into 10 statistically independent blocks for 
estimating errors. In addition, we use the method of histogram 
re-weighting~\cite{ferrenberg} to extrapolate for values of $\mu$ that 
are not directly simulated. This allows us to determine quantities like 
the maximum value of susceptibility and its location more precisely.

\section{\label{III}Two transitions in the 4-NN model}

\begin{figure}
\includegraphics[width=0.9 \columnwidth]{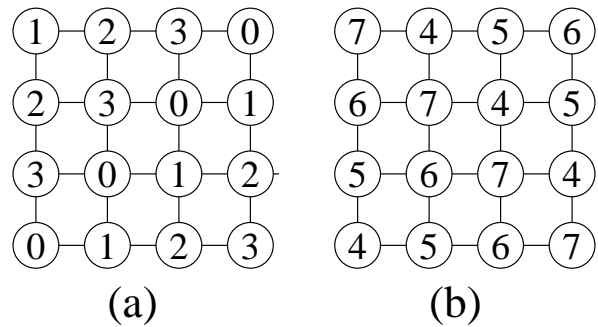}
\caption{The sublattice labeling for 4-NN. (a) All sites belonging to a 
diagonal oriented in the $\pi/4$ direction have same label. If the site 
has coordinates $(x,y)$, then the label is $[(x-y) \mod 4]$. (b) All 
sites belonging to a diagonal oriented in the $3 \pi/4$ direction have 
same label. If the site has coordinates $(x,y)$, then the label is 
$[(x+y) \mod 4 +4]$. In addition, all sites with label $0$ or $2$ (or 
equivalently $4$ and $6$) will be called sublattice $A$ and all sites 
with label $1$ or $3$ (or equivalently $5$ and $7$) will be called 
sublattice $B$. }
\label{fig:sublattice}
\end{figure}

In this section, we show numerically that the 4-NN model undergoes two phase 
transitions with increasing density. To assist in defining the different 
phases, we divide the lattice into sublattices by assigning $2$ labels 
to each site. Each lattice site belongs to a diagonal oriented in the 
$\pi/4$ direction and to a diagonal oriented in the $3 \pi/4$ direction. 
All sites that belong to a diagonal with orientation $\pi/4$ are 
assigned a label from $0$ to $3$ as shown in 
Fig.~\ref{fig:sublattice}(a). If the coordinates of a site are $(x,y)$, 
then the label is $[(x-y) \mod 4]$. Similarly, all sites that belong to 
a diagonal with orientation $3 \pi/4$ are assigned a label from $4$ to 
$7$ as shown in Fig.~\ref{fig:sublattice}(b). If the coordinates of a 
site are $(x,y)$, then the label is $[(x+y) \mod 4 +4]$.
\begin{figure}
\includegraphics[width=\columnwidth]{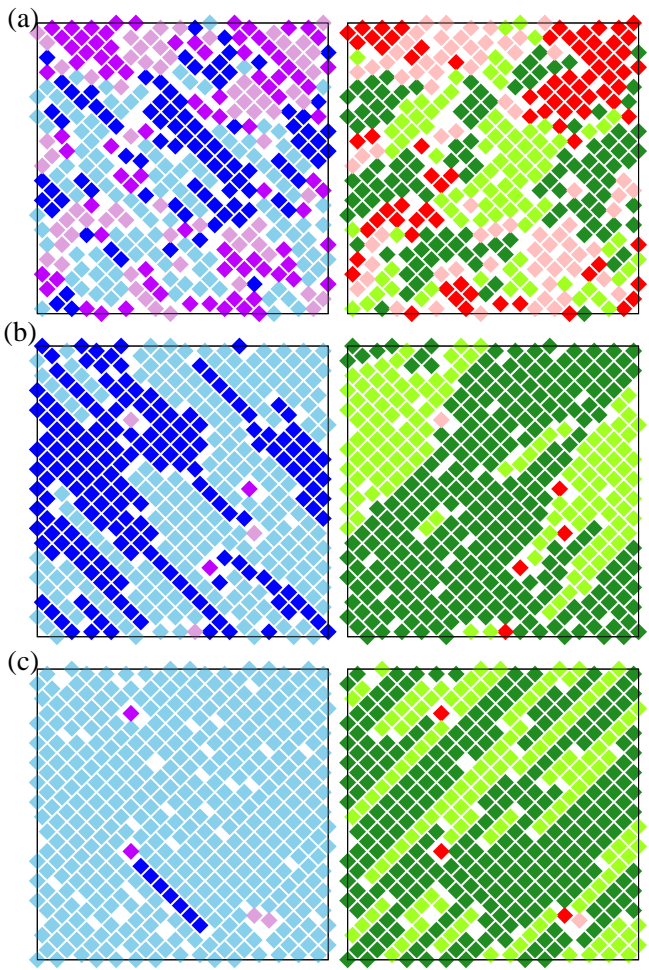}
\caption{Snapshots of typical configurations of the 4-NN model. The particles 
are colored according 
to the sublattice it belongs to (see Fig.~\ref{fig:sublattice}). In the 
left panel, the colors are light blue for 0, dark magenta for 1, deep 
blue for 2 and plum for 3. In the right panel, the colors are dark 
green for 4, red for 5, light green for 6 and pink for 7. (a) Low 
density disordered phase where all four colors are present in both 
panels. (b) Intermediate density sublattice phase, where even or odd 
sublattices are preferentially occupied. (c) High density columnar phase, 
where one sublattice in one of the panels (left in figure) and two sublattices 
in the other panel are preferentially occupied.
}
\label{fig:snapshots}
\end{figure}

In Fig.~\ref{fig:snapshots}, we show typical snapshots of the 
equilibrated system at low, intermediate and high densities. In the left 
panels, all particles belonging to sublattice $i$ ($i=0,1,2,3$) have the 
same color. Similarly, in the right panels, all particles belonging to 
sublattice $i$ ($i=4,5,6,7$) have the same color. At low densities [see 
Fig.~\ref{fig:snapshots}(a)], all four colors are present (roughly 
equal) in both the left and right panels. This is the disordered phase 
with equal occupation of all sublattices $0,\ldots,7$. At intermediate 
densities [see Fig.~\ref{fig:snapshots}(b)], we find that that majority 
of particles have two of the four colors in both left and right panels. 
This corresponds to particles preferably occupying either the even 
sublattices or the odd sublattices. There are two such states 
corresponding to particles in sublattices ($0, 2$) and ($4, 6$) or in 
sublattices ($1, 3$) and ($5, 7$). If we label the sites on sublattices 
$0$ and $2$ (equivalently $4$ and $6$) as $A$ and the sites on 
sublattices $1$ and $3$ (equivalently $5$ and $7$) as $B$, then this 
intermediate phase breaks the symmetry between $A$ and $B$ sublattices. 
We will call this phase a sublattice phase (following the terminology in 
Ref.~\cite{fernandes}). We note that this phase was observed in 
Ref.~\cite{fernandes}. At high densities [see 
Fig.~\ref{fig:snapshots}(c)], we find that the particles occupy one of 
the four sublattices from $0$--$3$ or $4$--$7$, but not from both. In 
the example shown in Fig.~\ref{fig:snapshots}(c), in the left panel, 
particles preferentially occupy sublattice $0$. However, in the right 
panel, particles occupy mostly two sublattices ($4$ and $6$). This phase 
is identical to the high density phase of the hard square model (2-NN) 
where there is positional order in one direction but no positional order 
in the perpendicular direction due to a sliding instability. We call 
this phase the columnar phase. There are $8$ such states, corresponding 
to the number of sublattices.

To distinguish between phases quantitatively, we define two order 
parameters $Q_{sl}$ and $Q_{cl}$, where $sl$ denotes sublattice and $cl$ 
denotes columnar. Let $\rho_i$, $i=0,\ldots,7$ be the density of 
particles in sublattice $i$. We define
\begin{subequations}\label{Q1}
\begin{eqnarray}
Q_{sl} & =&   \lvert(\rho_0 + \rho_2) - (\rho_1 + \rho_3)  \rvert,\label{Q1a}\\
Q_{cl} & =&  \lvert \sum_{k=0}^3 \rho_k e^{i k\pi/2}  \rvert - 
 \lvert \sum_{k=4}^7 \rho_k e^{i k \pi/2}\rvert. \label{Q1b}
\end{eqnarray}   
\end{subequations}
$Q_{sl}$ measures the difference between the densities of sublattices $A$ (sites 
in even sublattices) and $B$ (sites in odd sublattices). It is zero in 
the disordered phase and non-zero in both the sublattice and columnar 
phases. $Q_{cl}$ is zero in both the disordered and sublattice phases 
and non-zero only in the columnar phase.
\begin{figure}
\includegraphics[width=\columnwidth]{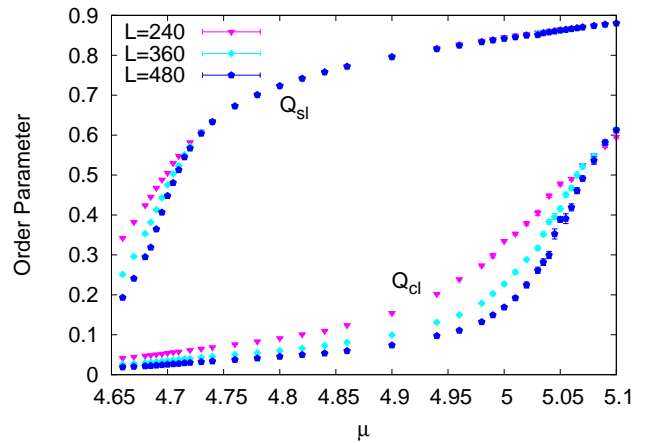}
\caption{(Color online) The variation of the order parameters $Q_{sl}$ and $Q_{cl}$ with 
chemical potential $\mu$ for different system sizes $L$. $Q_{sl}$ becomes
non-zero at a smaller value of $\mu$ than $Q_{cl}$.
}
\label{fig:Q_total}
\end{figure}

While the snapshots in Fig.~\ref{fig:snapshots} are indicative of two 
transitions, we now show unambiguously the existence of two transitions. 
Figure~\ref{fig:Q_total} shows the variation of the two order parameters 
with chemical potential. Clearly, $Q_{sl}$ takes on a $L$ independent 
non-zero value when the chemical potential is larger than $\mu\approx 
4.70$ ($\rho\approx0.110$). At this value of $\mu$ $Q_{cl}$ is still 
zero. $Q_{cl}$ takes on a $L$ independent non-zero value when the 
chemical potential is larger than $\mu \approx 5.07$ 
($\rho\approx0.116$). The two values of $\mu$ being clearly different, 
we conclude that there are two transitions.

The other thermodynamic quantities of interest are the susceptibility 
$\chi$, the second moment of the order parameter $\chi^{(2)}$, the 
Binder cumulant $U$, and compressibility $\kappa$, defined as
\begin{subequations}\label{cumulant}
\begin{eqnarray}
\chi& = &L^2[\langle Q^2\rangle-\langle Q\rangle^2],\label{cumulantchi}\\
\chi^{(2)}& = &L^2\langle Q\rangle^2,\label{cumulantq2}\\
U & = &1-\frac{\langle Q^4\rangle}{3\langle Q^2\rangle^2},\label{cumulantu}\\
\kappa& = &L^2[\langle \rho^2\rangle-\langle
\rho\rangle^2],\label{cumulantkappa}
\end{eqnarray}
\end{subequations}
where Q represent Q$_{sl}$ or Q$_{cl}$. Though $\chi$ and $\chi^{(2)}$ resemble
each other, we find that the data for $\chi^{(2)}$ is much cleaner.
The nature of the phase transitions 
is determined by the singular behavior of $U$, $Q$, $\chi$, and $\chi^{(2)}$
near the critical point. Let $\epsilon = (\mu-\mu_c)/\mu_c$, where
$\mu_c$ is the critical chemical potential. The singular behavior is
characterized by the critical exponents $\nu,$ $\beta$, $\gamma$, and $\alpha$
defined by $Q \sim (-\epsilon)^\beta$, $\epsilon<0$, $\chi \sim
|\epsilon|^{-\gamma}$, $\chi^{(2)} \sim
|\epsilon|^{-\gamma}$, $\kappa \sim
|\epsilon|^{-\alpha}$and $\xi \sim |\epsilon|^{-\nu}$,
where $\xi$ is the correlation length and $|\epsilon| \rightarrow 0$. 
The other critical exponents may be obtained from scaling relations. The 
exponents are obtained by finite size scaling of the different 
quantities near the critical point:
\begin{subequations}\label{scale}
\begin{eqnarray}
U&\simeq& f_U(\epsilon L^{1/\nu}),\label{scaleu}\\
Q &\simeq& L^{-\beta/\nu}f_Q(\epsilon L^{1/\nu}),\label{scaleq}\\
\chi &\simeq& L^{\gamma/\nu}f_{\chi}(\epsilon 
L^{1/\nu}),\label{scaleq2}\\
\chi^{(2)}&\simeq& L^{\gamma/\nu}f^{(2)}_{\chi}(\epsilon 
L^{1/\nu})\label{scalechi},\\
\kappa &\simeq& L^{\alpha/\nu}f_{\kappa}(\epsilon 
L^{1/\nu}),\label{scalekappa}
\end{eqnarray}
\end{subequations}
where $f_U$, $f_Q$, $f_{\chi}$, $f^{(2)}_{\chi}$, $f_\kappa$ are scaling 
functions and the system size is $L\times L$. In addition, if $\chi$ has 
a maximum $\chi_{max}(L)$ at $\mu_c(L)$, then
\begin{subequations}\label{muscale}
\begin{eqnarray}
\chi_{max}(L)&\propto&  L^{\gamma/\nu},\label{muscalea}\\
\mu_c(L)-\mu_c(\infty)&\propto& L^{-1/\nu}\label{muscaleb}.
\end{eqnarray}
\end{subequations}
\begin{figure}
\includegraphics[width=\columnwidth]{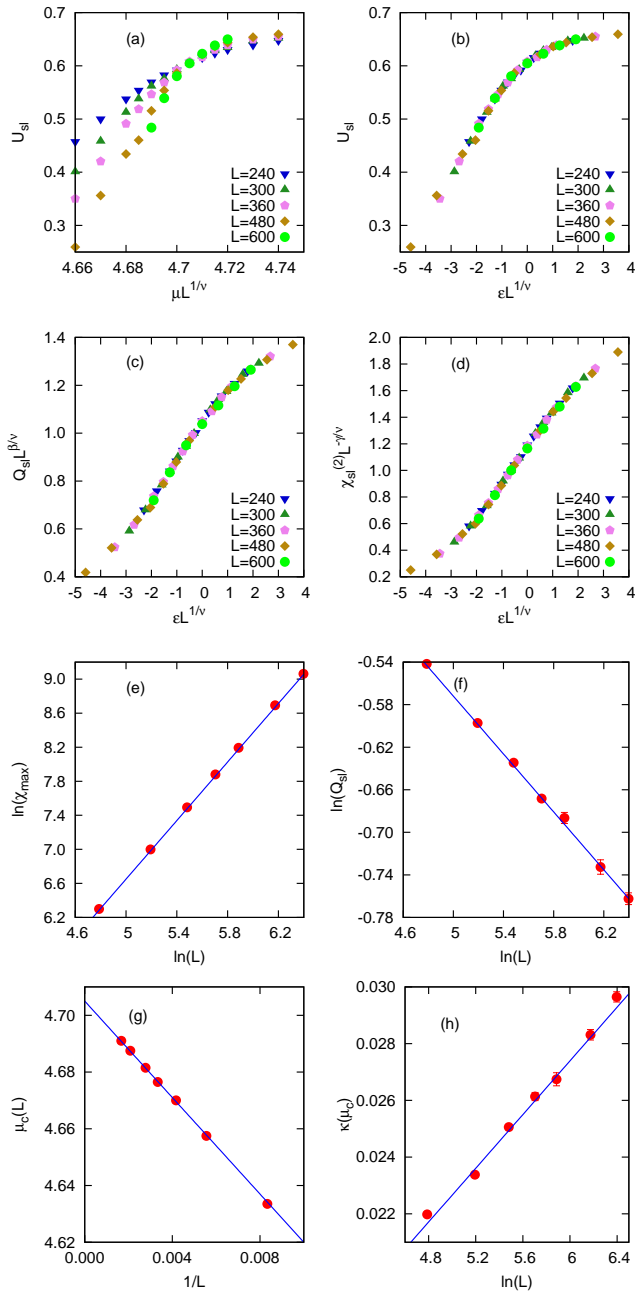}
\caption{(Color online) The data for cumulants of $Q_{sl}$ for the 4-NN model near 
the first transition from the disordered phase to sublattice phase. (a) The 
Binder cumulant $U_{sl}$ for different L intersect at $\mu_c=4.705$. The 
data for (b) $U_{sl}$, (c) $Q_{sl}$, and (d) $\chi_{sl}^{(2)}$ for 
different $L$ collapse onto a single curve when scaled as in Eq.~(\ref{scale})
with the Ising exponents $\beta/\nu =1/8$, $\gamma/\nu =7/4$, and 
$\nu=1$. (e) The variation of the maximum value $\chi_{max}$ of the 
susceptibility $\chi$ with $L$. The solid line is a best fit to the data 
with slope $\gamma/\nu=1.72 \pm 0.04$.  (f) The variation of the 
order parameter $Q_{sl}$ at the critical point with $L$. The solid line 
is a best fit to the data with slope $-\beta/\nu=-0.136 \pm 0.017$. 
(g) Variation of $\mu_c$ for different $L$ with $1/L$. The straight line
intersects the $\mu$-axis at $\mu_c=4.705\pm 0.005$. 
(h)  The variation of the compressibility  $\kappa$ at the critical
point with $L$. The straight line shows $\kappa(\mu_c) \sim \ln L$.
}
\label{fig:proof_sl}
\end{figure}

The first transition from disordered to sublattice phase was studied in 
Ref.~\cite{fernandes} using system sizes varying from $L=80$ to
$L=240$, and was argued to be in the Ising universality class. However,
analysis of the cluster integrals predict non-Ising
exponents~\cite{rotman}.
We re-examine this transition with data for system sizes up to $L=600$. In 
addition to obtaining better estimates of the critical exponents, it 
also acts as a check for our Monte Carlo algorithm. The data for the 
different thermodynamic quantities near the disordered-sublattice 
transition are shown in Fig.~\ref{fig:proof_sl}. The critical chemical 
potential $\mu_c$ is obtained from the intersection of the curves for the Binder 
cumulant $U_{sl}$ for different system sizes. The intersection 
point depends very weakly on $L$ [see Fig.~\ref{fig:proof_sl}(a)] 
allowing for an accurate determination of $\mu_c$. We thus obtain 
$\mu_c=4.705\pm0.005$, consistent with the $\mu_c$ found in 
Ref.~\cite{fernandes}. In the disordered-sublattice transition, the 
system breaks the symmetry between $A$ and $B$ sublattices. Due to the 
two-fold symmetry, we expect this transition to be in universality class 
of the two dimensional Ising model. Indeed, we find excellent data 
collapse when the data for $U_{sl}$ [see Fig.~\ref{fig:proof_sl}(b)], 
$Q_{sl}$ [see Fig.~\ref{fig:proof_sl}(c)] and $\chi_{sl}^{(2)}$ [see
Fig.~\ref{fig:proof_sl}(d)] are scaled as in Eq.~(\ref{scale}) with Ising exponents 
$\beta/\nu=1/8$, $\gamma/\nu=7/4$ and $\nu=1$.

We also estimate $\beta/\nu$ and $\gamma/\nu$ independently. The maximum 
value of susceptibility $\chi_{max}$ scales with $L$ as $L^{\gamma/\nu}$. We 
calculate $\chi_{max}$ by the method of histogram re-weighting, and 
obtain $\gamma/\nu=1.72 \pm 0.04$ [Fig.~\ref{fig:proof_sl}(e)]. The 
order parameter $Q_{sl}$ at the critical point decreases with $L$ as 
$L^{-\beta/\nu}$. By simulating for different $L$, we obtain 
$\beta/\nu=0.136\pm 0.017$ [see Fig.~\ref{fig:proof_sl}(f)]. Both these 
numerical values are consistent with the Ising exponents. 
In Fig.~\ref{fig:proof_sl}(g),
we show the variation of  $\mu_c(L)$ with $1/L$. The data lie on a straight line,
consistent with $\nu=1$. The intersection of the straight line 
with the $\mu$-axis gives $\mu_c=4.705 \pm 0.005$, consistent with the estimation
from the crossing of the curves for the Binder cumulant. Finally, we examine the data
for compressibility $\kappa$ at the critical point. The data for compressibility is more
noisy that that for other quantities, but is consistent with a logarithmic divergence with
$L$ [see Fig.~\ref{fig:proof_sl}(h)], as expected for the Ising universality class. Thus, we 
conclude, as in Ref.~\cite{fernandes} and contrary to the conclusion in 
Ref.~\cite{rotman}, that the transition from disordered to sublattice
phase is in the Ising universality class, 

We now focus on the second transition from the sublattice phase to the 
columnar phase. Suppose, in the sublattice phase, the system is in 
sublattice $A$. This corresponds to all sites belonging to sublattices 
$0,2,4,6$. In the second transition, the system picks out one of the 
four sublattices, with equal occupation of two other sublattices. Since 
this transition breaks a four fold symmetry, we expect this transition, if continuous,
to be in the universality class of the two color Ashkin Teller model. We, thus, expect 
$\gamma/\nu=7/4$, $\beta/\nu=1/8$, and $\nu$ depending on the parameters 
of the problem at hand~\cite{baxterBook}.  We now provide numerical evidence of the same.
\begin{figure}
\includegraphics[width=\columnwidth]{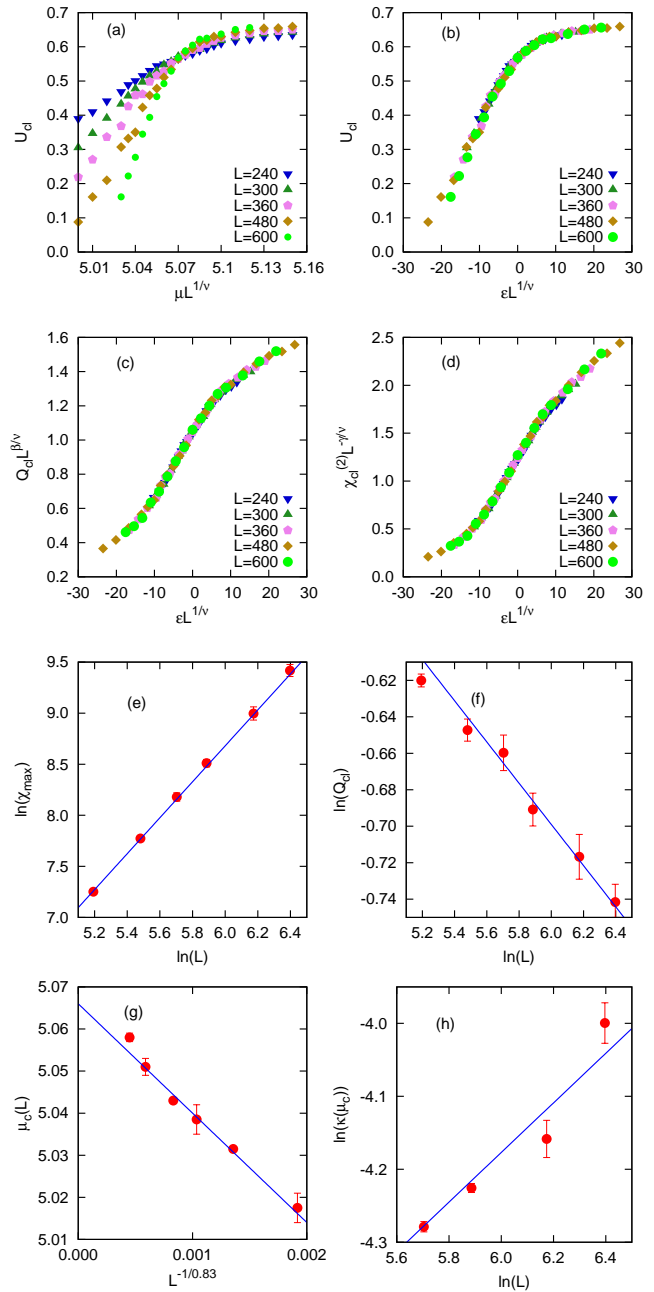}
\caption{(Color online) The data for cumulants of $Q_{cl}$ for the 4-NN model near the 
second transition from the sublattice phase to the columnar phase. (a) 
The Binder cumulant $U_{cl}$ for different $L$ intersect at 
$\mu_c=5.07\pm 0.01$.  The data for (b) $U_{cl}$, (c) $Q_{sl}$ and (d) 
$\chi_{sl}^{(2)}$ for different $L$ collapse onto a single curve when 
scaled as in Eq.~(\ref{scale}) with critical exponents $\beta/\nu =1/8$, 
$\gamma/\nu=7/4$ and $\nu=0.83\pm 0.06$. (e) The variation of the 
maximum value $\chi_{max}$ of the susceptibility $\chi$ with $L$. The 
solid line is a best fit to the data with slope $\gamma/\nu=1.76\pm 
0.05$. (f)The variation of the order parameter $Q_{sl}$ at the critical 
point with $L$. The solid line is a best fit to the data with slope 
$-\beta/\nu=-0.113\pm0.015$. (g) Variation of $\mu_c$ for different $L$ with $L^{-1/\nu}$.
The straight line intersects $\mu$-axis at $\mu_c=5.066 \pm 0.01$.
(h) The variation of compressibility $\kappa$ at the critical point $\mu_c$ with $L$.
The solid line is a best fit to the data with slope 
$\alpha/\nu=-0.34\pm 0.08$.
}
\label{fig:proof_cl}
\end{figure}

The data for the different thermodynamic quantities for the second 
transition are shown in Fig.~\ref{fig:proof_cl}. From the intersection 
of the curves of $U_{cl}$ for different $L$ [see 
Fig.~\ref{fig:proof_cl}(a)], we obtain $\mu_c=5.07\pm0.01$. The Binder 
cumulant data for different system sizes collapse onto a single curve when scaled 
as in Eq.~(\ref{scaleu}) with $\nu=0.083 \pm 0.06$ [see Fig.~\ref{fig:proof_cl}(b)]. 
For this value of 
$\nu$ and $\gamma/\nu=7/4$, $\beta/\nu=1/8$, we obtain excellent data 
collapse for $Q_{sl}$ [see Fig.~\ref{fig:proof_cl}(c)] and 
$\chi_{sl}^{(2)}$ [see Fig.~\ref{fig:proof_cl}(d)] when scaled as in Eq.~(\ref{scaleq})
and Eq.~(\ref{scaleq2}) respectively. Thus, we conclude 
that the transition belongs to the Ashkin Teller universality class with 
$\nu=0.83 \pm 0.06$, lying between the $4$ state Potts and Ising points. 
Independent measurement of $\gamma/\nu$ from the variation of the 
maximum of susceptibility with $L$ gives $\gamma/\nu=1.76 \pm 0.05$ [see 
Fig.~\ref{fig:proof_cl}(e)]. Similarly, from the dependence of $Q_{sl}$ 
on $L$ at the critical point, we obtain $\beta/\nu=0.113\pm 0.015$ [see 
Fig.~\ref{fig:proof_cl}(f)]. Both these values are consistent with the 
Ashkin Teller universality class.
In Fig.~~\ref{fig:proof_cl}(g), we show the dependence
of the critical chemical potential $\mu_c(L)$ on $L$. When plotted against $L^{-1/\nu}$
[see Eq.~(\ref{muscaleb})], with $\nu=0.83$, the data lie on a straight line which intersects
the $\mu$-axis at $\mu_c=5.066 \pm 0.01$. This estimate of the critical $\mu$ is 
consistent with the value obtained from the intersection of the curves for the Binder
cumulants.
Since $\nu<1$, the exponent $\alpha>0$ 
and we expect the compressibility $\kappa$ to diverge at the critical 
point with exponent $\alpha/\nu$. The data for $\kappa$ is very noisy 
[see Fig.~\ref{fig:proof_cl}(h)] when compared to data for other 
thermodynamic quantities. Fitting to a power law, we obtain  $\alpha/\nu=0.34\pm 
0.08$ . Within error bars, $\alpha/\nu$ 
satisfies the exponent equality $2 \nu= 2-\alpha$.

We conclude that the 4-NN model, contrary to what was known earlier,
undergoes two entropy driven transitions with increasing density. In order
to provide some understanding of this phenomena, we derive the high
density expansion of the model in the next section.

\section{\label{IV}High density expansion of the 4-NN model}

In this section, we calculate the first four terms in the large $z$
(high density) expansion of the free energy for the 4-NN model. We
show that the high density phase (say all particles in mostly sublattice $0$) 
has a sliding instability only for defects in sublattice $2$ and not 
for defects in sublattices $1$ and $3$. This results in the densities of
defects
in the different sublattices being different at large $z$. We argue that this sliding
instability could be
the origin of the two phase transitions in the 4-NN model.

The high density phase of the 4-NN model being columnar, the high
density expansion is very similar to that for the 2-NN
model~\cite{bellemans_nigam1nn,ramola}. 
Due to the sliding instability, the large $z$ expansion of the 2-NN
model is in powers of $1/\sqrt{z}$ instead of the usual $1/z$ Mayer
series.  The first three  terms in the high density
expansion for the 2-NN model was obtained in
Ref.~\cite{bellemans_nigam1nn}.
More recently, it has been systematically extended to $4$
terms~\cite{ramola}. We will closely follow the calculations of
Ref.~\cite{ramola}, modifying when necessary for the 4-NN model.

Consider a fully packed configuration of the 4-NN model. It has
density $1/8$. If
the particles occupy one of the sublattice from $0$ to $3$, then they
occupy two sublattices from  $4$ to $7$ and vice-versa (see
Fig.~\ref{fig:sublattice} for labeling of sublattices). For example, if
the particles are all in sublattice $0$, then they are also
simultaneously in sublattices $4$ and $6$.
It is easy to see that the number of 
fully packed configurations is $8 (2^{L/2}-1)$, where we assume that $L$ 
is even. Though the degeneracy diverges with $L$, the entropy per unit
site is zero in the thermodynamic limit.

For constructing the large $z$ expansion, we will describe the lattice
sites only in terms of sublattices $0$ to $3$. Let the activities on
sublattice $i$ be $z_i$. We will consider $z_0 \gg z_i$, $i=1,2,3$.
After the expansion is obtained, we will equate all the activities to
$z$. Thus, it will be an expansion about an ordered state where all
the particles are in sublattice $0$. The free energy
$f(z_0,z_1,z_2,z_3)$ is defined as
\begin{equation}\label{free1}
f(z_0,z_1,z_2,z_3)=\lim_{N\rightarrow \infty} \frac{-1}{N} 
\ln\mathcal{L}(z_0,z_1,z_2,z_3),
\end{equation}
where $\mathcal{L}(z_0,z_1,z_2,z_3)$ is the grand canonical partition 
function. To lowest order, the contribution to the partition function is 
$\mathcal{L}(z_0,0,0,0)$. It is easy to see that the partition function 
breaks up into a product of one dimensional partition functions, and is
\be
\mathcal{L}(z_0,0,0,0)= \Omega_p(z_0,L)^{L/4}.
\ee
Here $\Omega_p(z,\ell)$ and $\Omega_o(z,\ell)$ are the partition 
functions of a nearest neighbor exclusion hard core gas on a one 
dimensional lattice of length $\ell$ with periodic and open boundary 
conditions respectively, as defined in Eq.~(\ref{omega}) with $d=1$.

Solving Eq.~(\ref{omega}), we obtain~\cite{bellemans_nigam1nn,ramola}
\bea
\Omega_o(z,\ell) &= & \frac{\lambda_+^{\ell+2} -
\lambda_-^{\ell+2}}{\sqrt{1+4 z}}, ~\ell=0,1,\ldots,\\
\Omega_p(z,\ell) &= & \lambda_+^{\ell} +
\lambda_-^{\ell}, ~\ell=2,3,\ldots,
\eea
where
\be
\lambda_\pm = \frac{1 \pm \sqrt{1+4 z}}{2}.
\ee
Knowing the partition functions, the contribution from configurations
with zero defects (all particles in
sublattice $0$) to the free energy is
\be
f(z_0,0,0,0) = \frac{-\ln z_0}{8} - \frac{1}{8 \sqrt{z_0}} +
\frac{1}{192 z_0^{3/2}} + O(\frac{1}{z_0^{5/2}}).
\label{eq:nodefect}
\ee

\begin{figure}
\includegraphics[width=\columnwidth]{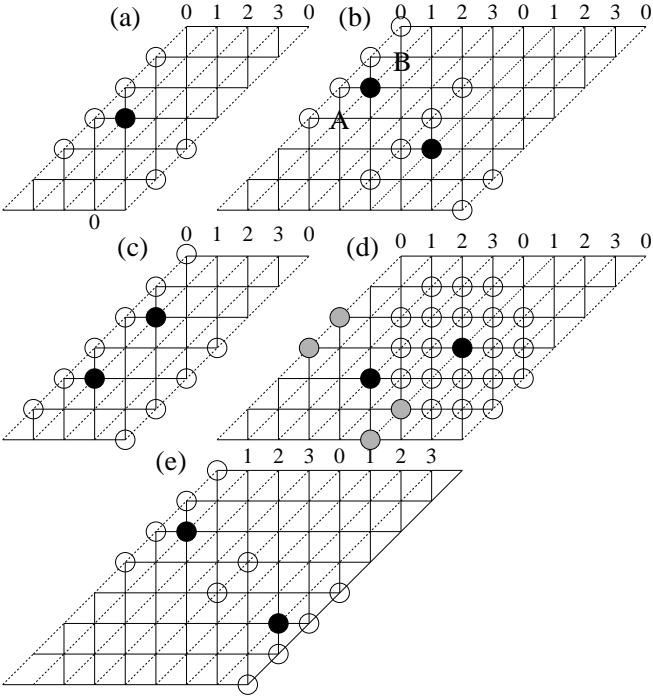}
\caption{Examples of one and two defect configurations on sublattices
$1$ and $3$ for an ordered state where particles are on sublattice $0$. 
Black circles are particles, empty and lightly shaded
circles are excluded sites. Exclusions
only on sublattice $0$ are shown in (a)--(c) and (e). 
$0,1,2,3$ denote the diagonals that belong
to sublattices $0,1,2,3$. (a) A single defect on
sublattice $1$.   (b) Two defects on
sublattice $1$ but on neighboring  diagonals. A and B are two 
other lattice sites where the defect may be placed. (c) Two defects on 
sublattice $1$ but on the same
diagonal. (d) A defect on 
sublattice $1$ and a defect on
sublattice $3$ at one of the $4$ closest positions. The empty circles
are all the excluded sites due to defect on sublattice $1$ while the
lightly shaded circles are excluded sites on sublattice $0$ due to
defect on sublattice $3$. (e) A defect on sublattice $1$ and a defect on
sublattice $3$ positioned such that they exclude the same two sites
on the  diagonal $0$ separating them.
}
\label{fig:fig09}
\end{figure}

We now switch on $z_1$. This creates some defect sites in sublattice
$1$. We first calculate the contribution from configurations with a single defect. A
single defect on sublattice $1$ excludes $4$ sites from the 
diagonal of sublattice $0$ closest to it and 2 sites from the diagonal 
further away from it [see Fig.~\ref{fig:fig09}(a)]. The
contribution of configurations with one defect to the partition function is
\be
\left. \frac{\mathcal{L}(z_0,z_1,0,0)}{\mathcal{L}(z_0,0,0,0)}
\right|_{1D} = \frac{N z_1}{4} \frac{\Omega_o(z_0,L-4) \Omega_o(z_0, L-2)}
{\Omega_p(z_0,L)^{2}},
\label{eq:ratio}
\ee
where the factor $N/4$ is the number of ways of placing a particle on
sublattice $1$, and $1D$ denotes one defect. Expanding for large $z_0$, 
we obtain
\be
\left. \frac{\mathcal{L}(z_0,z_1,0,0)}{\mathcal{L}(z_0,0,0,0)}
\right|_{1D} = N z_1\left[\frac{1}{16z_0^2} - \frac{1}{16 z_0^{5/2} }+
O(z_0^{-3})
\right].
\label{eq:1defect}
\ee

We now consider the contribution from configurations with two defects on sublattice $1$.
The lowest order contributions come from the two defects being on the
same diagonal or on adjacent diagonals. First consider two defects on adjacent diagonals,
as shown in Fig.~\ref{fig:fig09}(b). Now, $4$, $4$ and $2$ lattice sites
are excluded from diagonals belonging to sublattice $0$. The ratio of
the partition functions with $z_1\neq0$ and $z_1=0$ 
for this configuration with two defects is
\be
\frac{3N z_1^2}{4} \frac{\Omega_o(z_0,L-4)^2 \Omega_o(z_0,L-2)}
{\Omega_p(z_0,L)^3},
\ee
where the factor $3 N/4$ is the combinatorial factor associated with the
number of ways of placing the pair of particles. For each choice of the position
of the first particle ($N/4$ ways), there are 3 ways of placing the second particle
(A, B, and filled circle). Expanding for large
$z_0$, we obtain
\be
N z_1^2\left[\frac{3}{32 z_0^{7/2}} + O(z_0^{-4}) \right].
\ee

Second, consider the case when 
two defects are on sublattice $1$ but on the same diagonal, as shown
in Fig.~\ref{fig:fig09}(c). Now,
$6$ and $4$ lattice sites are excluded from diagonals belonging to
sublattice $0$. The ratio of 
the partition functions for this two defect
configuration is
\be
\frac{N z_1^2}{4} \frac{\Omega_o(z_0,L-6) \Omega_o(z_0,L-4)}
{\Omega_p(z_0,L)^2},
\ee
where the factor $N/4$ is the combinatorial factor associated with the
number of ways of placing the pair of particles. Expanding for large 
$z_0$, we obtain
\be
N z_1^2 O(z_0^{-4}).
\ee
Thus, to order $z^{-3/2}$, there is no contribution.
Collecting together the terms, we obtain
\be
\left. \frac{\mathcal{L}(z_0,z_1,0,0)}{\mathcal{L}(z_0,0,0,0)}
\right|_{2D} = N z_1^2 \left[\frac{3}{32 z_0^{7/2}}+ O(z_0^{-3}) \right].
\label{eq:2defect}
\ee

It is straightforward to verify that the contribution from
configurations with $3$ defects do not contribute to terms up to
order $z^{-3/2}$.

We now switch on a small $z_3$.
Sublattices $1$ and $3$ being symmetric with respect to sublattice
$0$, the contribution from configurations with a single defect on sublattice $3$ is
identical to Eq.~(\ref{eq:1defect}) except for $z_1\to z_3$.
Similarly, configurations with two defects on sublattice $3$ have the same contribution as
Eq.~(\ref{eq:2defect}) with $z_1\to z_3$. We now calculate the
contribution to the partition function from configurations with one defect on sublattice
$1$ and another on sublattice $3$ as shown in
Fig.~\ref{fig:fig09}(d) and (e). In Fig.~\ref{fig:fig09}(d), the particles are placed as
close to each other as possible. Given a particle on sublattice $1$
(placed in $N/4$ ways), there are four ways of placing a particle on
sublattice $3$.
The ratio of
the partition functions for this two defect
configuration is
\be
N z_1 z_3 \frac{\Omega_o(z_0,L-2)^2 \Omega_o(z_0,L-6)}
{\Omega_p(z_0,L)^3}.
\ee
Expanding for large
$z_0$, we obtain
\be
N z_1 z_3\left[\frac{1}{8 z_0^{7/2}} + O(z_0^{-4}) \right].
\ee

Now, consider the configuration shown In Fig.~\ref{fig:fig09}(e).
Once the first particle is placed (in $N/4$ ways), there is a unique position
for the second particle.  The ratio of partition functions  
when two such defects are present is
\be
\frac{N z_1 z_3 }{4}\frac{\Omega_o(z_0,L-4)^2 \Omega_o(z_0,L-2)}
{\Omega_p(z_0,L)^3}.
\ee
Expanding for large
$z_0$, we obtain
\be
N z_1 z_3\left[\frac{1}{32 z_0^{7/2}} + O(z_0^{-4}) \right].
\ee

Combining together the contributions from configurations with  one or two defects
on sublattices
$1$ and $3$, we obtain
\bea\label{12d}
&&\frac{\mathcal{L}(z_0,z_1,0,z_3)}{\mathcal{L}(z_0,0,0,0)}
= N (z_1+z_3) \left[\frac{1}{16z_0^2} - \frac{1}{16 z_0^{5/2}
} \right]\nonumber\\
&& + \frac{3N(z_1^2+z_3^2)}{32 z_0^{7/2}} 
+ \frac{5 N z_1 z_3 }{32 z_0^{7/2}} + O(z^{-2}).
\eea

\begin{figure}
\includegraphics[width=0.9 \columnwidth]{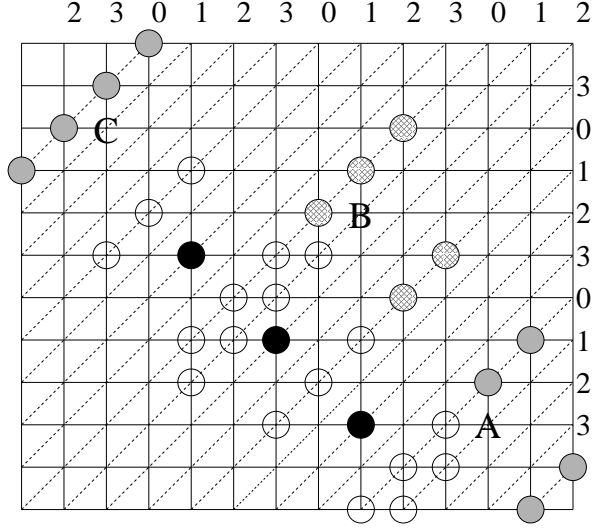}
\caption{Example of a rod-like defect on sublattice $2$ and defects
on sublattice $1$.
Black circles are particles on sublattice $2$. Empty circles are
excluded sites on sublattice $0$ or sublattice $1$ (all not shown)
due to defect on sublattice $2$.
Lightly shaded circles are excluded sites  due to
defects at $A$ or $C$. Circles with
a pattern are excluded sites  due to defect at $B$.}
\label{fig:fig10}
\end{figure}

We now focus on defects in sublattice $2$. Unlike defects in
sublattices $1$ and $3$, a vacancy on sublattice $0$ can be broken into
two half vacancies with the points in between being defects on
sublattice $2$. Thus, $n$ defects on sublattice $2$
contribute at the same order as a single defect~\cite{ramola}. 
The expansion is better performed in terms of rods
which are a collection of contiguous defects in the $3 \pi/4$
direction on sublattice $2$~\cite{ramola}. An example of a rod of
length $3$ is shown in Fig.~\ref{fig:fig10} (focus only on black
circles 
and empty circles that are triplets). It excludes three
sites each from $4$ diagonals belonging to sublattice $0$. It is
straightforward to
obtain the contribution from a single rod~\cite{ramola}:
\begin{subequations}\label{1d1r}
\begin{eqnarray}
\left. \frac{\mathcal{L}(z_0,0,z_2,0)}{\mathcal{L}(z_0,0,0,0)}
\right|_{1R}& =& \frac{N }{4}\sum_{n=1}^\infty
z_2^n \left[\frac{\Omega_o(z_0,L-3)}
{\Omega_p(z_0,L)} \right]^{n+1},\label{1d1ra}\\
&=& \frac{N}{4} \frac{z_2 \beta^2}{1-z_2 \beta}\label{1d1rb},
\end{eqnarray}
\end{subequations}
where $1R$ denoted one rod and 
\be
\beta= \frac{1}{\sqrt{1+4 z_0} \lambda_+},
\ee
is a function only of $z_0$.

The calculation of the contribution from two rods on sublattice $2$ is
identical to that for the 2-NN model, except for an overall factor of $1/2$ due
to only half the lattice sites being involved in the calculation.
Thus, one can read off the results from the calculations of
Ref.~\cite{ramola} for the 2-NN model. 

A further contribution to the free energy at order $z^{-3/2}$ is due
to configurations with a rod on sublattice $2$ and a particle on sublattice $1$ or $3$.
Given a rod  on sublattice $2$, there are three kinds of
sites on sublattice $1$ where a particle may be placed. These are
denoted by $A$, $B$ and $C$ (see Fig.~\ref{fig:fig10}). The
contribution to the ratios of partition functions
$\mathcal{L}(z_0,z_1,z_2,0)/\mathcal{L}(z_0,0,0,0)$ for the different
cases are
\bea
A&:&\frac{N}{4} 
z_2^n \left[\frac{\Omega_o(z_0,L-3)} {\Omega_p(z_0,L)} \right]^{n}
\nonumber \\
&& \times 2 z_1 \frac{\Omega_o(z_0,L-5)\Omega_o(z_0,L-2) }
{\Omega_p(z_0,L)^2} \sim O(z^{-3/2}),\\
B&:&\frac{N}{4} z_2^n 
\left[\frac{\Omega_o(z_0,L-3)} {\Omega_p(z_0,L)} \right]^{n-1}
\nonumber \\
&& \times 2 n z_1 \frac{\Omega_o(z_0,L-6) \Omega_o(z_0,L-5)}
{\Omega_p(z_0,L)^2} \sim O(z^{-5/2}),\\
C&:&\frac{N}{4} z_2^n 
\left[\frac{\Omega_o(z_0,L-3)} {\Omega_p(z_0,L)} \right]^{n}
\nonumber \\
&& \times 2 z_1 \frac{\Omega_o(z_0,L-4)}
{\Omega_p(z_0,L)} \sim O(z^{-3/2}),
\eea
where $n$ is the length of the rod.
Thus to order $z^{-3/2}$, only cases $A$ and $C$ are relevant, 
and their contribution to the free energy add up to 
\bea\label{1R_1D}
\left. \frac{\mathcal{L}(z_0,z_1,z_2,z_3)}{\mathcal{L}(z_0,0,0,0)}
\right|_{1R,1D}=
\frac{N}{4} \frac{z_2 \beta}{1-z_2 \beta} \frac{(z_1+z_3)}{2 z_0^{5/2}}\nonumber \\
+ \frac{N}{4} \frac{z_2 \beta^2}{1-z_2 \beta} \frac{(z_1+z_3)}{ z_0^{3/2}}+O(z^{-2}),
\eea
where, in the right hand side of Eq.~(\ref{1R_1D}), the first term is due to $A$ and 
the second term is due to $C$.

The free energy of the 4-NN model may now be written down up to
$O(z^{-3/2})$. In terms of the partition functions,
\be
f(z_0,z_1,z_2,z_3) = -\frac{\ln \mathcal{L}(z_0,0,0,0)}{N}- \frac{
\mathcal{L}(z_0,z_1,z_2,z_3)}{N  \mathcal{L}(z_0,0,0,0)}.
\ee
Adding the contribution from configurations with two rods, as obtained
in
Ref.~\cite{ramola}, to  Eqs.~(\ref{eq:nodefect}), (\ref{12d}), (\ref{1d1rb}),
and (\ref{1R_1D}), and equating $z_i=z$, we obtain
\begin{equation}\label{free2}
-f(z)=\frac{1}{8} \ln z+\frac{1}{8z^{1/2}}+\frac{1}{4z}+
\frac{\frac{3}{2}\ln \frac{9}{8}+\frac{149}{192}}{z^{3/2}}+ O(z^{-2}).
\end{equation}

Knowing the free energy, the particle densities in each sublattice is
given by $\rho_i=z_i \partial/\partial z_i (-f)$. Doing the algebra
and simplifying,
\bea
\rho(z) & = & \frac{1}{8} -\frac{1}{16 z^{1/2}}-\frac{1}{4z}-
\frac{\frac{9}{4}\ln \frac{9}{8}+\frac{149}{128}}{z^{3/2}}+
O(\frac{1}{z^2}),
\label{eq:rho}\\
\rho_0(z) & = & \frac{1}{8} -\frac{1}{16 z^{1/2}}-\frac{5}{8z}-
\frac{\frac{605}{128}+\frac{17}{4}\ln \frac{9}{8}}{z^{3/2}}+
O(\frac{1}{z^2}),
\label{eq:rho0}\\
\rho_1(z) & = & \frac{1}{16 z}+
\frac{17}{32 z^{3/2}}+ O(\frac{1}{z^2}),
\label{eq:rho1}\\
\rho_2(z) & = & \frac{1}{4z}+
\frac{2\ln \frac{9}{8}+\frac{5}{2}}{z^{3/2}}+
O(\frac{1}{z^2}).
\label{eq:rho2}
\eea
\begin{figure}
\includegraphics[width=\columnwidth]{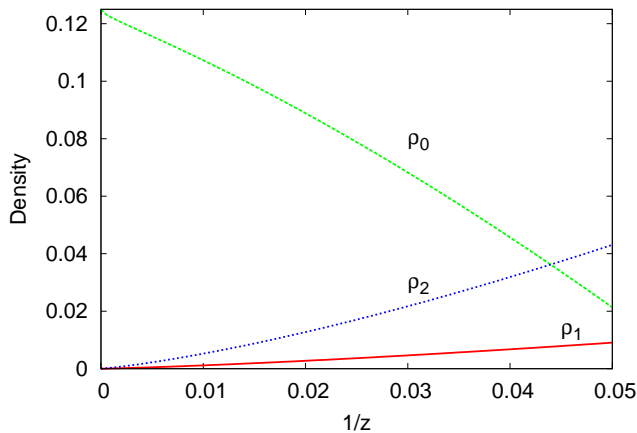}
\caption{(Color online) Variation of the densities of particles in different sublattices, 
truncated at order $z^{-3/2}$ [see Eqs.~(\ref{eq:rho0}),
(\ref{eq:rho1}), (\ref{eq:rho2})], with $1/z$.}
\label{fig:fig11}
\end{figure}

The sublattice densities, truncated at order $z^{-3/2}$
are plotted in Fig.~\ref{fig:fig11}.
Clearly, due to the sliding instability, the density on sublattice $2$
increases faster than that of sublattices $1$ and $3$. Equating
$\rho_0$ and $\rho_2$ in Eqs.~(\ref{eq:rho0}) and (\ref{eq:rho2}), 
an estimate of the
critical activity may be estimated. We find $z_c=22.742\ldots$ or
$\mu_c = 3.1242\ldots$. This should be compared with the actual value
of $\mu_c \approx 5.07$.

The faster increase in the particle density of the sublattice where
sliding instability exists is the likely reason for the second
transition. As density is increased, the system is first destabilized
by the sliding instability into sublattices $0$ and $2$. This is followed by
a second transition where the density of sublattices $1$, $3$ equals that of 
$0$, $2$.

\section{\label{V} Multiple transitions in the $k$-NN HCLG}

In this section, we generalize the arguments of Sec.~\ref{IV} to
larger $k$. We ask for a criteria that will help determine whether the
HCLG for a given $k$ will undergo multiple transitions. From the
analysis of the 4-NN model, it is clear that if the high density phase is
columnar, then it is easier to generate defects in the sublattices
where the sliding instability is present as compared to other
sublattices. Hence, we conjecture that if the model satisfies (i) the
high density phase is columnar and (ii) the sliding instability is
present in only a fraction of the sublattices, then the system will
show multiple transitions.
\begin{figure}
\includegraphics[width=\columnwidth]{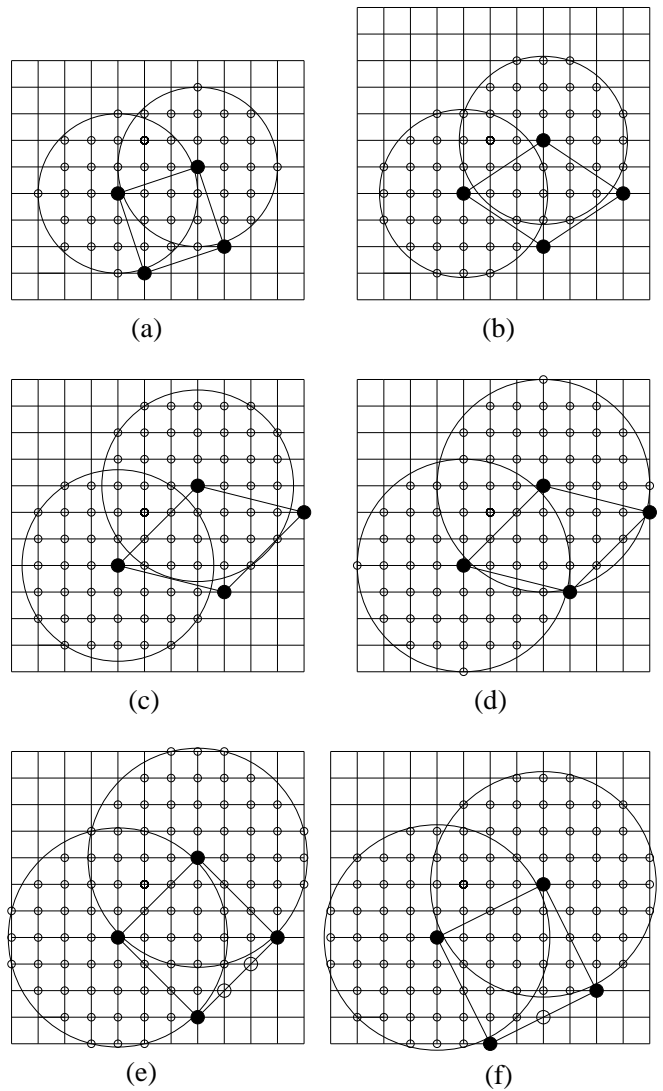}
\caption{ 
One of the configurations of maximum density for different $k$.
Particles are shown by large filled circles. The excluded sites are
shown by small circles. A
circle that encloses the excluded sites due to a particle are drawn
around the left most and top most particles. The empty large circles denote sites that may be
occupied by sliding all particles on that diagonal, keeping other
particles fixed. The examples are for 
(a) 6-NN, 
(b) 7-NN, 
(c) 8-NN, 
(d) 9-NN,
(e) 10-NN, and  
(f) 11-NN. 
}
\label{fig:fig12}
\end{figure}

In hard square systems ($k=2, 5$), though
the high density phase is columnar, the  
sliding instability is along all sublattices and hence does not
satisfy condition (ii). Hence, one expects only one transition for the
hard square system, at least for $2\times 2$ and $3\times 3$ systems.
Models, other than hard square models, that show columnar order at
high densities will typically have sliding instability in only a few
sublattices. For $k>5$, none of the $k$-NN HCLG models are 
hard square models. Thus, the
criteria reduces to determining whether the high density phase is
columnar or not.

The nature of the phase at high densities may be determined by
constructing configurations at full packing for different $k$. In
Fig.~\ref{fig:fig12}, we show configurations at full packing for $k=6$
to $11$. In each of the plots, $4$ particles (filled circles) that
constitute the unit cell are shown. A circle of radius $R$, which
depends on $k$,  is drawn
around two of these particles --
the left most and top most. $R$ is distance of the farthest 
exclusion point from the particle hence all
lattice sites within or on the circle are excluded by the
particle. The value of $R$ for different $k$ is given in Table.~\ref{table:generalk}.  
In all the cases, the configuration can be thought of as
particles placed on equidistant parallel diagonals (not necessarily oriented
in the $\pi/4$ direction). In some cases [see Fig.~\ref{fig:fig12} (e)
and (f)], all the particles in a diagonal may be slid by one or two
lattice spacings without affecting the configurations in other
diagonals. Such allowed sites are denoted by empty large circles.
If such a freedom to slide exists,  the system will have columnar
order at high densities.
\begin{table}
\caption{\label{table:generalk} For each $k$, the square of radius of circle of exclusion $R^2$, 
density at full packing $\rho_{max}$  and the nature of the high density phase are tabulated.}
\begin{ruledtabular}
\begin{tabular}{cllc}
k & $R^2$ & $\rho_{max}$ & High density phase\\ \hline
   1 &  1 & 1/2 & Sublattice \\
   2 &  2 & 1/4 & Columnar \\
   3 &  4 & 1/5 & Sublattice \\ 
   4 &  5 & 1/8 & Columnar \\ 
   5 &  8 & 1/9 & Columnar \\ 
   6 &  9 & 1/10 & Sublattice \\ 
   7 &  10 & 1/12 & Sublattice \\ 
   8 &  13 & 1/15 & Sublattice \\ 
   9 &  16 & 1/15 & Sublattice \\ 
  10 &  17 & 1/18 & Columnar\\ 
  11 & 	18 & 1/20 & Columnar \\
  12 &  20 & 1/23 & Sublattice \\
  13 &  25 & 1/24 & Sublattice \\
  14 &  26 & 1/28 & Columnar
  \end{tabular}
\end{ruledtabular}
\end{table}

For 6-NN, in the fully packed configuration, the particles are along
diagonals oriented in the $\tan^{-1}(1/3)$ or
$\tan^{-1}(3)$ directions [see Fig.~\ref{fig:fig12}(a)]. There is no
freedom to slide, and hence we expect sublattice order at full
packing. This is true for $k=7,8,9$, where the diagonals are oriented
in different directions for different $k$ [see
Fig.~\ref{fig:fig12}(b)--(d)]. Thus, the conjecture would predict a
single first order transition to an ordered sublattice phase for
$k=6,\ldots,9$.

For the 10-NN model, the particles in any diagonal may be slid by one or two
lattice sites in the $\pi/4$ direction without affecting the
configurations in the other diagonals [see Fig.~\ref{fig:fig12}(e)]. For
the 11-NN model, all the particles in a diagonal may be slid by one
lattice spacing in the $\tan^{-1}(1/2)$ direction without affecting
the configurations in the other diagonals [see Fig.~\ref{fig:fig12}(f)].
Thus, the conjecture predicts that there should be multiple
transitions in the 10-NN and 11-NN model.

The analysis is easily extended to larger $k$ by constructing the
fully packed configurations. For instance, the next $k$ to have
columnar order at high densities is $k=14$. The nature of the high
density phase and the density at full packing for $k\leq 14$ are
summarized in Table.~\ref{table:generalk}. 

\section{\label{VI} Monte Carlo simulations for $k=6, \ldots, 11$}

In this section, we present results from Monte Carlo simulations for
$k$-NN HCLG models with $k=6,\ldots,11$ to verify the conjecture
presented in Sec.~\ref{V}. We refer to Sec.~\ref{II} and
Table~\ref{table:param} for details of the simulations. We present the
details for $k=6$ to $k=9$ in Sec.~\ref{6nn}, $k=10$ is
Sec.~\ref{10nn}, $k=11$ in Sec.~\ref{11nn}.

\subsection{The 6-NN to 9-NN  models\label{6nn}}

The conjecture in Sec.~\ref{V} predicts a single first order
transition from a disordered phase to an ordered sublattice phase for
$k=6,7,8,9$. In order to define suitable order parameters, we divide
the lattice sites into different sublattices. Like in the 4-NN model,
each site belongs to two sublattices.

\begin{figure}
\includegraphics[width=0.8 \columnwidth]{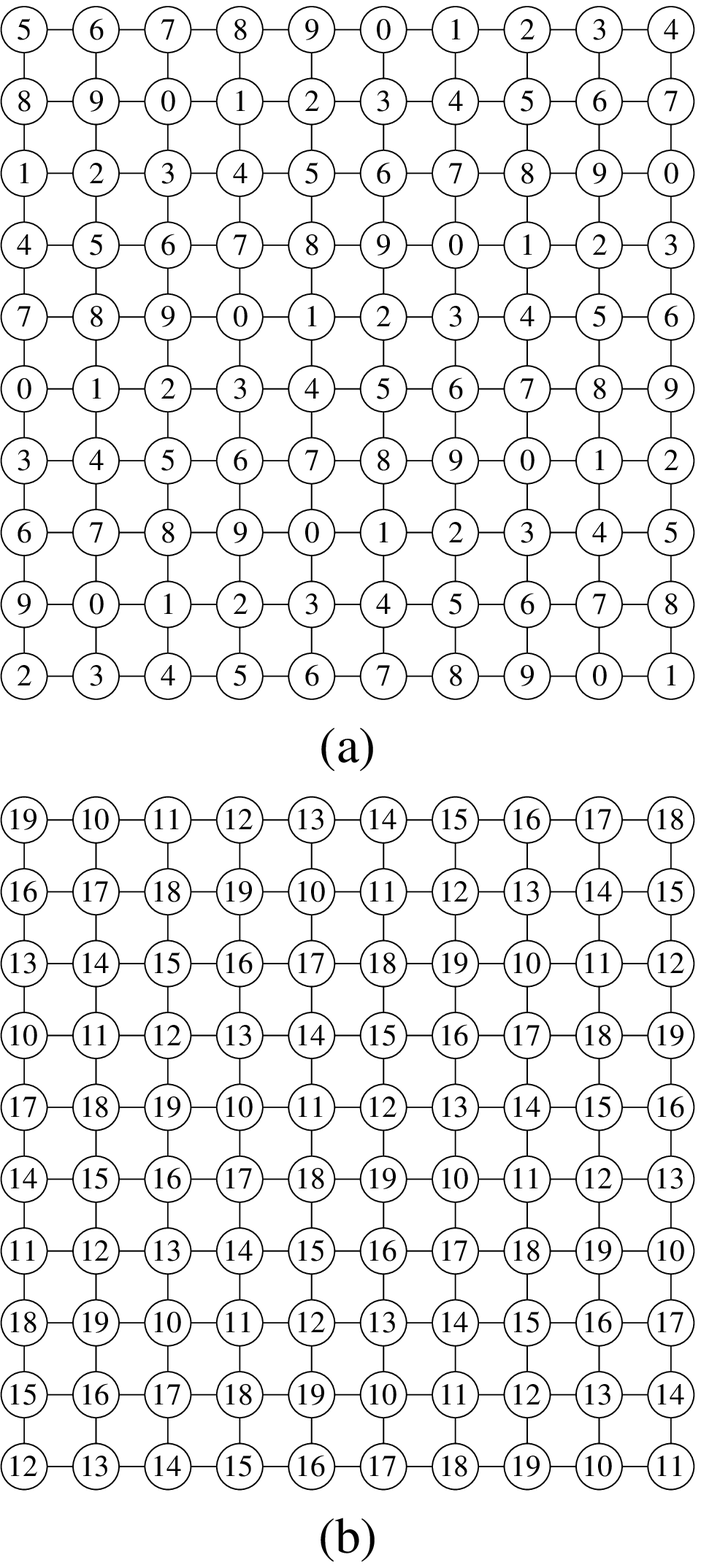}
\caption{The sublattices of the 6-NN model. (a) All sites on a diagonal 
oriented in the $\tan^{-1}(1/3)$ direction belong to the same
sublattice. (b) All sites on a diagonal 
oriented in the $\tan^{-1}(3)$ direction belong to the same 
sublattice.
}
\label{fig:fig13}
\end{figure}

The sublattices for the 6-NN model are shown in  Fig.~\ref{fig:fig13}. 
In Fig.~\ref{fig:fig13}(a) [Fig.~\ref{fig:fig13}(b)], 
all sites belonging to a diagonal oriented in the $\tan^{-1}(1/3)$ 
[$\tan^{-1}(3)$] direction belong to the same sublattice. There are
$10$ sublattices each for the two choices. In the high density phase,
most particles will occupy one of the $20$ sublattices with maximum density
$1/10$.

\begin{figure}
\includegraphics[width=\columnwidth]{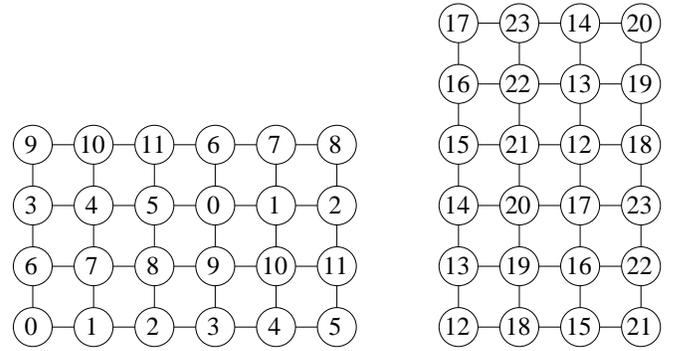}
\caption{The sublattices of the 7-NN model. (a) All sites on a diagonal 
oriented in the $\tan^{-1}(2/3)$ direction belong to the same
sublattice. (b) All sites on a diagonal 
oriented in the $\tan^{-1}(3/2)$ direction belong to the same 
sublattice.
}
\label{fig:fig14}
\end{figure}

The sublattices for the 7-NN model are shown 
in Fig.~\ref{fig:fig14}. 
In Fig.~\ref{fig:fig14}(a) [Fig.~\ref{fig:fig14}(b)], 
all sites belonging to a diagonal oriented in the $\tan^{-1}(2/3)$ 
[$\tan^{-1}(3/2)$] direction belong to the same sublattice. There are
$12$ sublattices each for the two choices. 
In the high density phase,
most particles will occupy one of the $24$ sublattices with maximum density
$1/12$.

\begin{figure}
\includegraphics[width=\columnwidth]{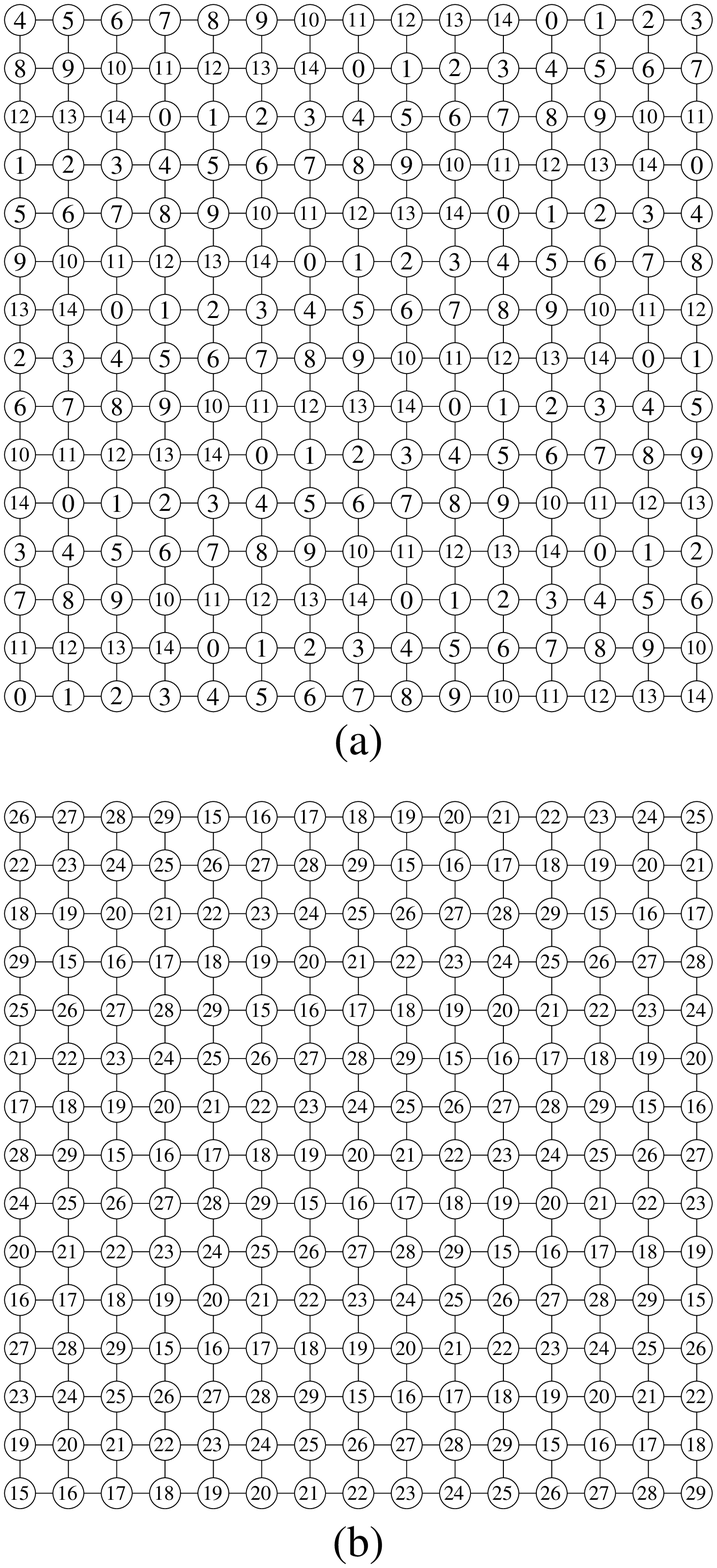}
\caption{The sublattices of the 8-NN and the 9-NN  models. (a) 
All sites on a diagonal 
oriented in the $\tan^{-1}(1/4)$ direction belong to the same
sublattice. (b) All sites on a diagonal 
oriented in the $\tan^{-1}(-4)$ direction belong to the same 
sublattice.
}
\label{fig:fig15}
\end{figure}

The sublattices for the 8-NN model are shown 
in Fig.~\ref{fig:fig15}. 
In Fig.~\ref{fig:fig15}(a) [Fig.~\ref{fig:fig15}(b)], 
all sites belonging to a diagonal oriented in the $\tan^{-1}(1/4)$ 
[$\tan^{-1}(-4)$] direction belong to the same sublattice. There are
$15$ sublattices each for the two choices. 
In the high density phase,
most particles will occupy one of the $30$ sublattices with maximum density
$1/15$.

The sublattices for the 9-NN model are identical to that for the 8-NN
model shown in Fig.~\ref{fig:fig15}, and hence maximum density will be
$1/15$.

To study the phase transition into the sublattice phase for $k=6$ to
$9$, we define an order parameter
\begin{equation}
Q_k = \lvert Q_k^{(a)}\rvert - \lvert Q_k^{(b)} \rvert,
\label{eq:orderparameter1}
\end{equation}
where $k=6,7,8,9$ denotes k-NN and $Q_k^{(a)}$ and $Q_k^{(b)}$ measure 
sublattice ordering according to the sublattice labeling in
(a) and (b) respectively of
Figs.~\ref{fig:fig13}, \ref{fig:fig14}, and \ref{fig:fig15} and are
defined as
\bea
Q_k^{(a)} &=&\sum_{j=0}^{m-1} \rho_j e^{2\pi \iota j/m},
\label{eq:orderparameter2} \\
Q_k^{(b)} &=&\sum_{j=m}^{2 m-1} \rho_j e^{2\pi \iota j/m}.
\label{eq:orderparameter3}
\eea
Here $\rho_j$ is the particle density in sublattice $j$. The parameter
$m$ depends on $k$ and has values $10$ (6-NN), $12$ (7-NN), $15$
(8-NN) and $15$ (9-NN).
Clearly,
$Q_i$ is zero in the disordered phase and non-zero in the sublattice
ordered phase.

We now study the transitions for $k=6$ to $k=9$ using the above order
parameter. To show that a transition is first order,
we measure the probability density function (pdf)
of the density $\rho$ and order parameter $Q_k$ near the transition for 
two different values of $L$. The pdfs should have two peaks that do
not move closer to each other with increasing $L$. One of the peaks
correspond to the disordered phase and the other to the ordered phase.
This will be taken as 
a signature of a first order transition.

We observe transitions at critical
chemical potentials $\mu_c\approx4.66$ ($k=6$), $\mu_c\approx4.88$
($k=7$), $\mu\approx6.07$ ($k=8$), and  $\mu\approx4.63$ ($k=9$).
Surprisingly, we find $\mu_c$ for 9-NN to be smaller than that for
6-NN. 
\begin{figure}
\includegraphics[width=\columnwidth]{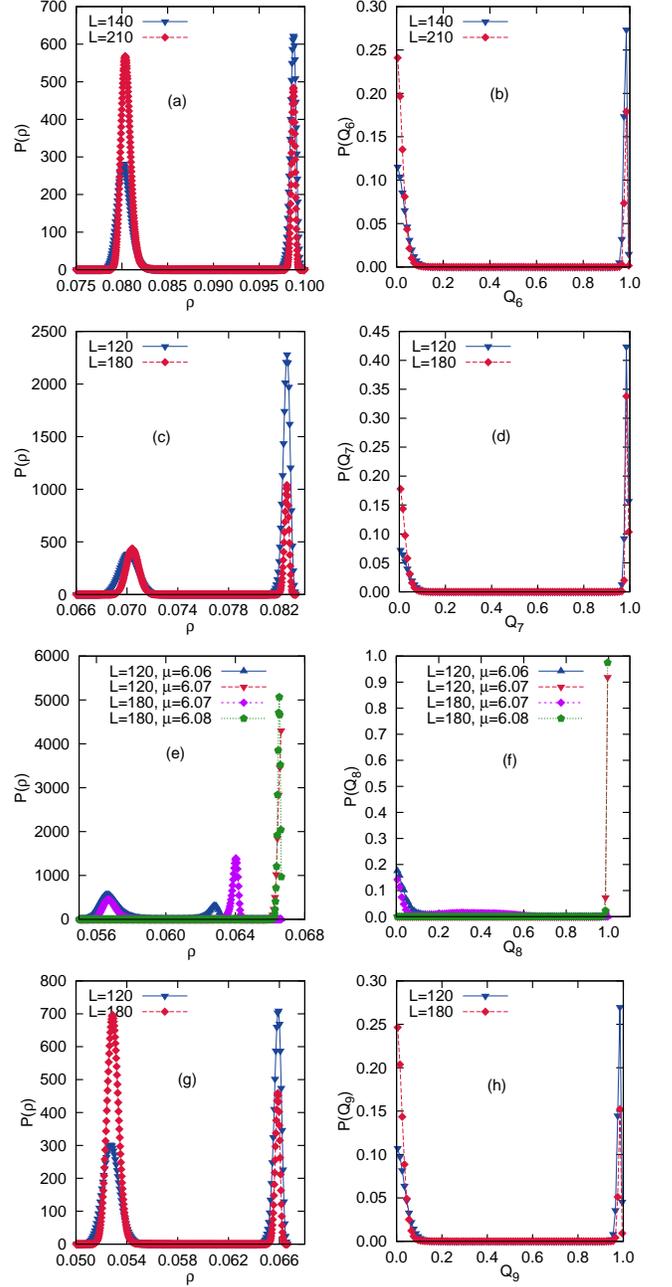}
\caption{(Color online) The probability density function for density $\rho$ (left
panels) and order parameter $Q_k$ (right panels) for $\mu \approx
\mu_c$, the critical chemical potential,
for two different system sizes. The data are
for the k-NN model with $k=6$ [(a) and (b)],
$k=7$ [(c) and (d)], $k=8$ [(e) and (f)], and $k=9$ [(g) and (h)].
}
\label{fig:fig16}
\end{figure}

The pdfs for $\rho$ and $Q_k$ are shown in Fig.~\ref{fig:fig16} (a),
(b) [6-NN], (c), (d) [7-NN], (e), (f) [8-NN] and (g), (h) [9-NN]. 
In all the plots, we observe that the pdfs have two peaks that do not
approach each with increasing system size. For 8-NN [see
Fig.~\ref{fig:fig16} (e), (f)], there is a long-lived metastable state
at a density that lies between those for the low-disordered phase and
the high-density sublattice phase. Hence, the pdf for $\mu>\mu_c$ is
peaked at a value different from the peak at the transition point.
We, therefore, conclude that the transitions in the 6-NN to 9-NN 
models are first
order.

We check that
the phase for values of $\mu \lesssim \mu_c$ is the disordered phase and for
values of $\mu \gtrsim \mu_c$ is the sublattice ordered phase (by
looking at typical snapshots)  expected at
full packing. Thus, we do not expect any more transitions. Both the first
order nature and the single transition are consistent with the
conjecture in Sec.~\ref{V}.

\subsection{The 10-NN model\label{10nn}}

\begin{figure}
\includegraphics[width= \columnwidth]{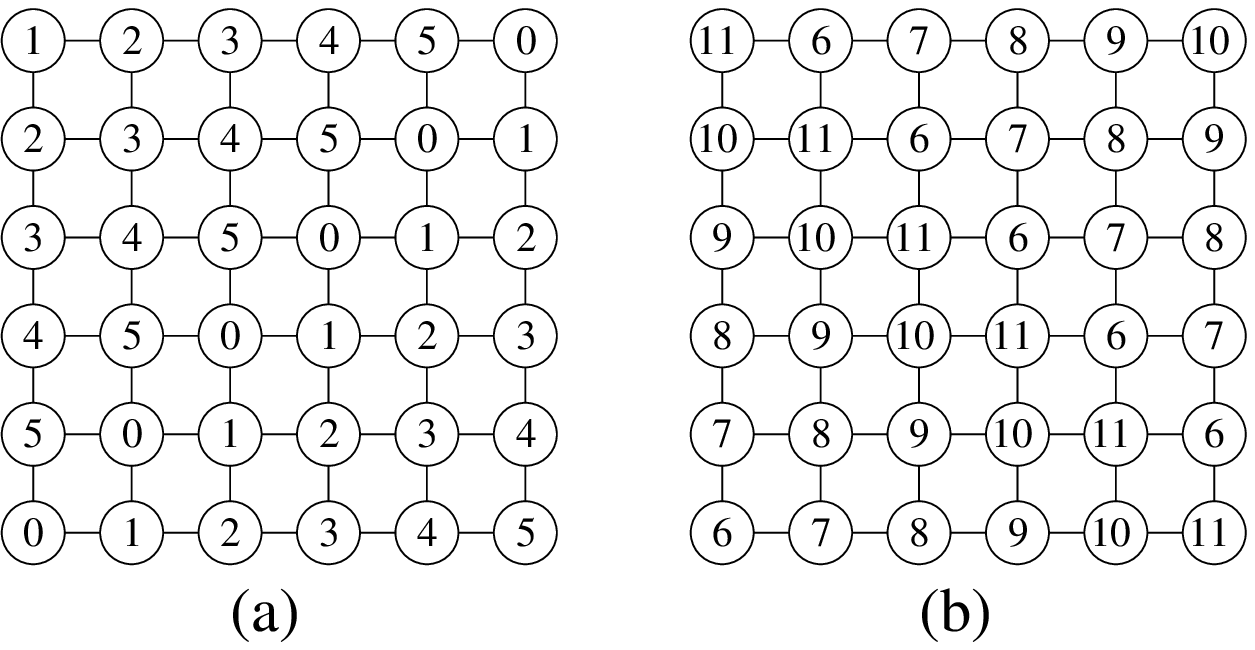}
\caption{The sublattices of the 10-NN model. (a) All sites on a diagonal 
oriented in the $\pi/4$ direction belong to the same
sublattice. (b) All sites on a diagonal 
oriented in the $3\pi/4$ direction belong to the same 
sublattice.
}
\label{fig:fig17}
\end{figure}

For the 10-NN model, the conjecture in Sec.~\ref{V} predicts multiple 
transitions. In this subsection, we confirm the same. 
We divide the lattice into sublattices as shown in Fig.~\ref{fig:fig17}. 
Each site belongs to two sublattices. In
Fig.~\ref{fig:fig17}(a) [Fig.~\ref{fig:fig17}(b)], 
all sites belonging to a diagonal oriented in the $\pi/4$ 
($3 \pi/4$) direction belong to the same sublattice. There are
$6$ sublattices each for the two choices. 

In the high density phase,
we expect the system to be in a columnar phase where all the 
particles occupy one sublattice either from $0$ to $5$ or from 
$6$ to $11$.  For example, if the particles are all in sublattice $0$ 
direction, they are also in sublattices $6$, $8$ and $10$.
The maximum density possible is $1/18$.

\begin{figure}
\includegraphics[width=0.9 \columnwidth]{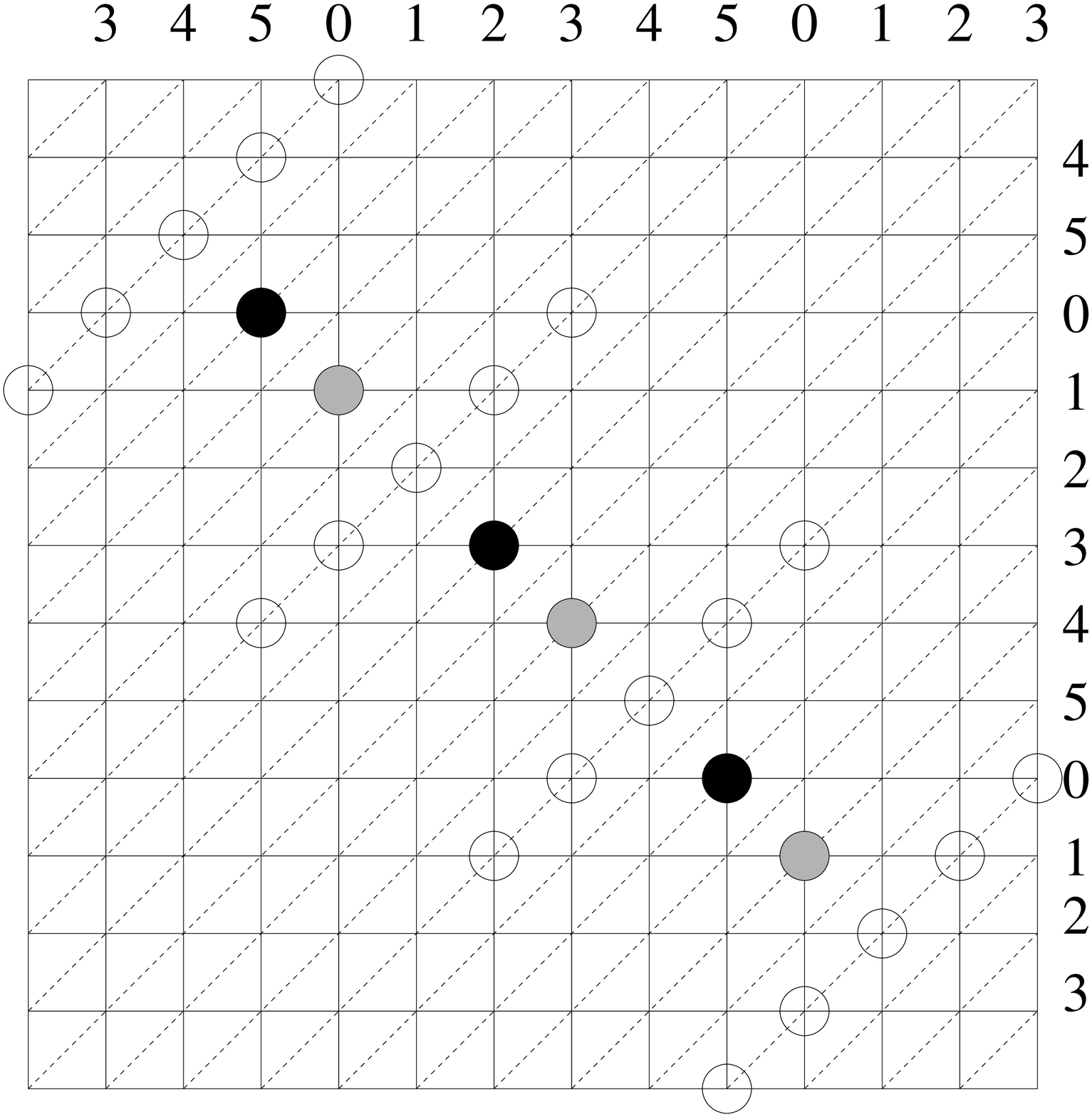}
\caption{An example of rod-like defects of length $3$ on sublattices $2$ (black circles) 
or $4$ (grey circles) in the 10-NN model.  Exclusions only on sublattice $0$  are shown by empty
circles. $0, 1, 2, 3, 4, 5$ on the edge of the box
denote diagonals that belong to sublattices $0,1,2,3,4,5$ respectively.
}
\label{fig:fig18}
\end{figure}

We first show that, if all the particles are in sublattice $0$ in the
columnar phase, then there is a sliding instability along sublattices
$2$ and $4$. To do so, we consider configurations with defects on
sublattice $2$ or sublattice $4$ that are rod-like, and show that rods
of all lengths contribute at same order. In Fig.~\ref{fig:fig18}, we
show examples of configurations with rods of length $3$ on sublattice
$2$ (black circles) and on sublattice $4$ (lightly shaded circles).
A rod of length $n$ excludes $5$ sites each from $n+1$ diagonals
belonging to sublattice $0$. The contribution from configurations with
rods of length $n$ to the ratio of partition functions 
$\mathcal{L}(z_0,0,z_2,0,0,0/\mathcal{L}(z_0,0,0,0,0,0)$
configuration is 
 \be
\frac{N }{6}
z_2^n  \left[\frac{\Omega_o(z_0,L-5)}
{\Omega_p(z_0,L)} \right]^{n+1}.
\label{eq:10nn1rod}
\ee
In the 10-NN model, along a diagonal, a particle excludes its nearest and next-nearest
neighbors from being occupied by a particle. Then, to leading order
$\Omega_o(z_0,L-5) \sim z_0^{L/3-1}$, and $\Omega_p(z_0,L) \sim  z_0^{L/3}$. Thus,
the term in Eq.~(\ref{eq:10nn1rod}) is $z^{-1}$ to leading order for all $n\geq 1$. 

Thus, the sliding instability exists along sublattices $2$ and $4$. It is straightforward to verify that
it does not exist on sublattices $1$, $3$, and $5$. The conjecture in Sec.~\ref{V} then predicts
that as $z$ is decreased, the columnar phase should first destabilize into a phase where 
particle densities will be equal on all even sublattices and all odd sublattices, but not 
equal to each other. If we label the even sublattices as $A$ and odd sublattices as $B$, then
there is a symmetry breaking between $A$ and $B$. Further decrease in $z$ would result
in a disordered phase. Like for the 4-NN model, we will call the intermediate phase as a 
sublattice phase.

In terms of increasing $z$ or $\mu$, we expect the first transition to be in the 
universality class of two dimensional Ising model because of symmetry breaking 
between two symmetric phases. In the second transition, the system
chooses from one of $6$ symmetric phases. By analogy with Potts model, we expect
the second transition to be first order. 

\begin{figure}
\includegraphics[width=\columnwidth]{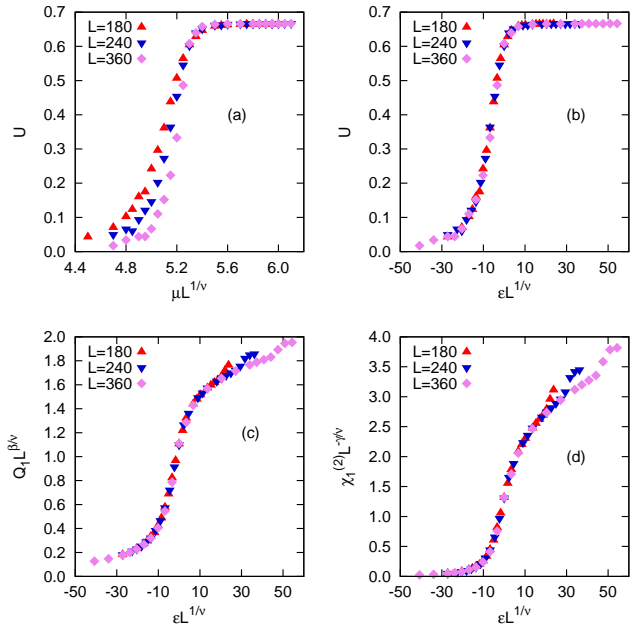}
\caption{(Color online) 
The data for cumulants of $Q_{1}$ for the 10-NN model near 
the first transition from the disordered phase to sublattice phase. (a) The 
Binder cumulant $U_{1}$ for different L crosses at $\mu_c=5.30$. The 
data for (b) $U_{1}$, (c) $Q_{1}$, and (d) $\chi_{1}^{(2)}$ for 
different $L$ collapse onto a single curve when scaled as in Eq.~(\ref{scale})
with the Ising exponents $\beta/\nu =1/8$, $\gamma/\nu =7/4$, and 
$\nu=1$.
}
\label{fig:fig19}
\end{figure}

We now confirm these predictions numerically.
To study the first transition, we define an order parameter,
\begin{equation}
Q_{1}  = \lvert(\rho_0 + \rho_2 +\rho_4) - (\rho_1 + \rho_3 + \rho_5)\rvert,
\label{Q5}
\end{equation}
where $\rho_i$ are the particle densities on sublattice $i$. 
$Q_1$ measures the density difference
between even and odd sublattices. 
The data for the 
different thermodynamic quantities near the disordered-sublattice 
transition are shown in Fig.~\ref{fig:fig19}. The critical chemical 
potential $\mu_c$ is obtained from the intersection of the Binder 
cumulant curves $U_1$ for different system sizes.  We  obtain 
$\mu_c=5.3\pm0.05$. The data for $U_{1}$ [see Fig.~\ref{fig:fig19}(b)], 
$Q_{1}$ [see Fig.~\ref{fig:fig19}(c)] and $\chi_{1}^{(2)}$ [see Fig.~\ref{fig:fig19}(d)] 
for different system sizes collapse onto a single curve
when scaled as  in Eq.~(\ref{scale}) with Ising exponents 
$\beta/\nu=1/8$, $\gamma/\nu=7/4$ and $\nu=1$.
These results confirm that the first transition is continuous and is consistent
to the Ising universality class.
\begin{figure}
\includegraphics[width=\columnwidth]{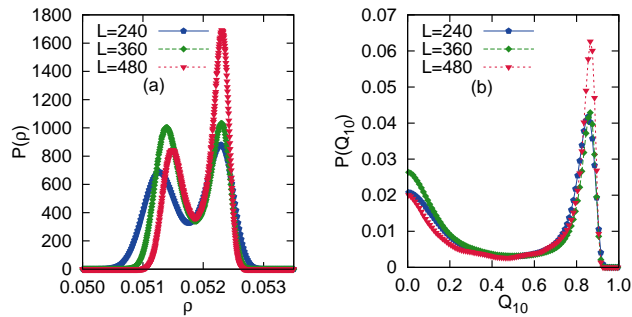}
\caption{(Color online) The probability density function for (a) density $\rho$ and (b)
order parameter $Q_{10}$ near the transition point
($\mu_c \approx 6.0$) for three different system sizes. The data are
for the 10-NN model.
}
\label{fig:fig20}
\end{figure}

We now study the second transition. This is best done using an
order parameter $Q_{10}$ as defined in 
Eqs.~(\ref{eq:orderparameter1}) and
(\ref{eq:orderparameter2}) with  $m=6$ in
Eq.~(\ref{eq:orderparameter2}). $Q_{10}$ is zero in disordered and sublattice phases and 
non-zero in the columnar phase. We find that $Q_{10}$ is zero for $\mu \lesssim 5.95$ and 
non zero for $\mu \gtrsim 6.00$. These values are distinctly larger than the critical
chemical potential found above for the first transition ($5.30$). To establish the first
order nature of the transition, we measure the pdfs of density $\rho$ and $Q_{10}$
near the transition point. These are shown in Fig.~\ref{fig:fig20} (a) and (b). The pdfs
for both quantities have two well separated peaks that become sharper with
increasing system size. This is a clear signature of a first order transition.

We check that
the phase  for
values of $\mu \gtrsim \mu_c$ is the columnar ordered phase (by
looking at typical snapshots)  expected at
full packing. Thus, we do not expect any more transitions. Thus, the numerical
data are consistent with the
conjecture in Sec.~\ref{V}.

\subsection{The 11-NN model \label{11nn}}

\begin{figure}
\includegraphics[width=\columnwidth]{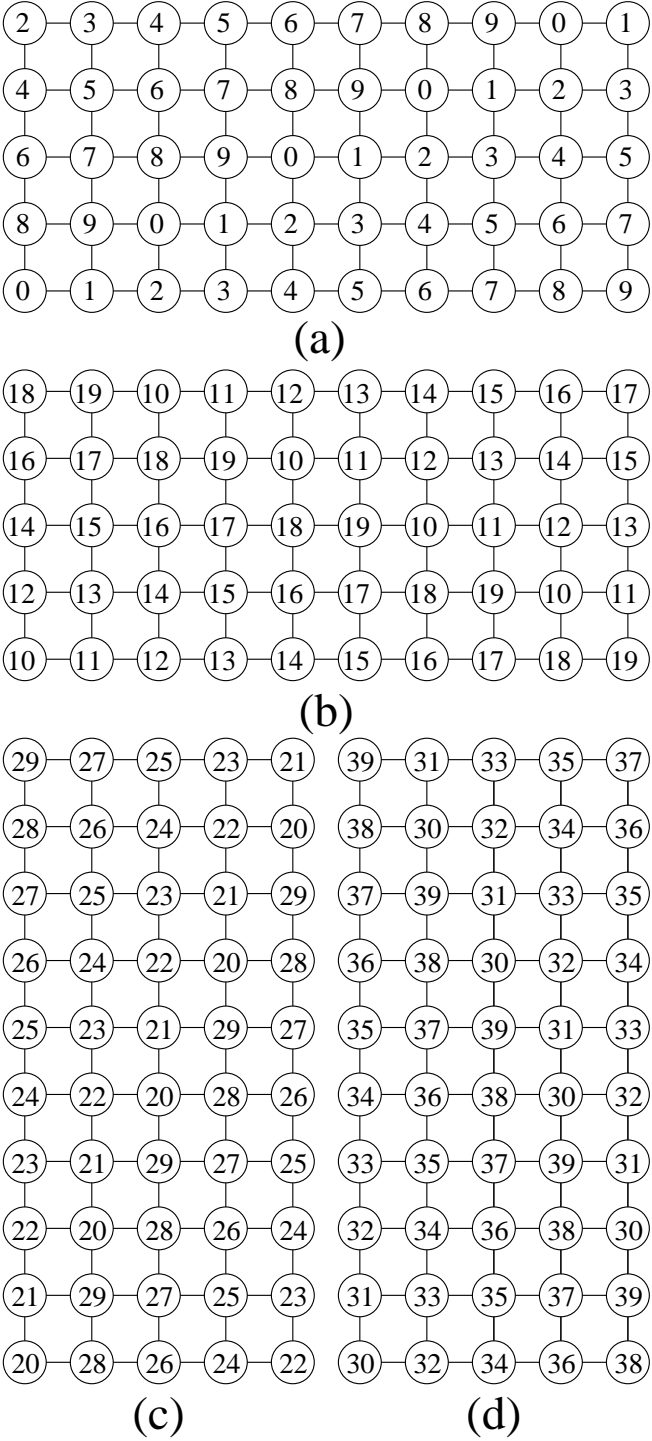}
\caption{The sublattices of the 11-NN model. (a) All sites on a diagonal 
oriented in the $\tan^{-1}(1/2)$ direction belong to the same
sublattice. (b) All sites on a diagonal 
oriented in the $\tan^{-1}(-1/2)$ direction belong to the same 
sublattice. (c) All sites on a diagonal 
oriented in the $\tan^{-1}(2)$ direction belong to the same
sublattice. (b) All sites on a diagonal 
oriented in the $\tan^{-1}(-2)$ direction belong to the same 
sublattice. 
}
\label{fig:fig21}
\end{figure}

For the 11-NN model, the conjecture in Sec.~\ref{V} predicts multiple 
transitions. In this subsection, we numerically confirm the same. 
We divide the lattice into sublattices as shown in Fig.~\ref{fig:fig21}. 
In contrast to sublattice decomposition for $k\leq 10$, now
each site belongs to four sublattices. In
Fig.~\ref{fig:fig17}(a), (b), (c), and (d),  
all sites belonging to a diagonal oriented in the 
$\tan^{-1}(1/2)$, $\tan^{-1}(-1/2)$, $\tan^{-1}(2)$, and 
$\tan^{-1}(-2)$ directions respectively
belong to the same sublattice. There are
$10$ sublattices  for each of the four choices. 

In the high density phase,
we expect the system to be in a columnar phase where all the 
particles occupy one sublattice chosen from one of the sets $0$ to $9$,
$10$ to $19$, $20$ to $29$, and $30$ to $39$.  An example
of a columnar phase would be all particles in sublattice $0$. In that
case, the particles are also 
simultaneously in all even sublattices between $10$ and $19$, all sublattices 
between $20$ and $29$, and sublattices $30$ and $35$. Along
a diagonal in sublattice $0$, there is nearest neighbor exclusion.
Hence, the  maximum density possible is $1/20$.

Consider, now, the high density phase. Suppose the ordered phase is
one in which all particles are present in sublattice $0$. We now
introduce defects and ask what sublattices have a sliding
instability. We skip the details, but it is straightforward to verify
that the sliding instability is present only for sublattice $5$. This
means that $n$-rod defects of  $n$ defects on sublattice $5$
contribute to the same order in the free energy. The conjecture in
Sec.~\ref{V} therefore predicts that the high density phase should
destabilize into a phase where two sublattices are equally occupied.
Further decrease in chemical potential would finally result in a
low density disordered phase.

We now present results from Monte Carlo simulations for the 11-NN
model. As chemical potential is increased, we find that the system
undergoes a first order transition (see below) at $\mu_c \approx
6.40$. However, for $\mu \gtrsim \mu_c$, the phase that we observe is
not the phase where $2$ sublattices are present. Instead, we observe
a phase in which particles are mostly present on 
even sublattices (for e.g., 0,2,4,6,8) or odd sublattices 
(for e.g.,1,3,5,7,9). Equivalently, in terms of rows and columns, the
particles occupy every fourth row or every fourth column. There are
$8$ such states. 

A convenient order parameter to study the transition is
\begin{equation}
Q_{11} = \lvert \sum_{j=0}^{3} r_j e^{2\pi \iota j/4} \rvert - 
\lvert \sum_{j=0}^{3} c_j e^{2\pi \iota j/4} \rvert 
\label{eq:orderparameter11}
\end{equation}
where $r_j$ is the particle density in rows $[j \mod 4]$ and $c_j$ is
the particle density in columns $[j \mod 4]$. $Q_{11}$ is zero in the
disordered phase and non-zero in the intermediate phase.

\begin{figure}
\includegraphics[width=\columnwidth]{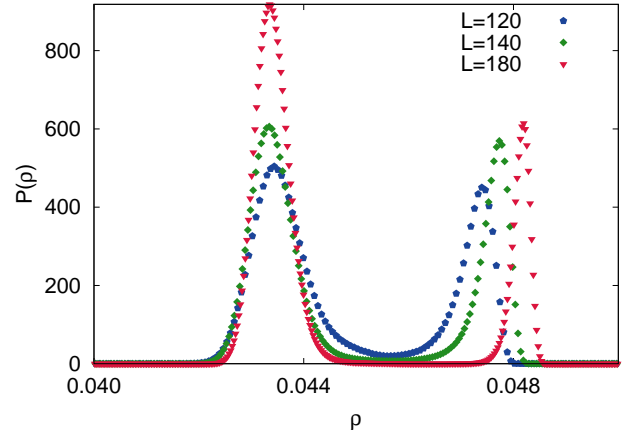}
\caption{(Color online) The probability distribution function for density
$\rho$ near the transition point for three different system sizes.
The data are for the 11-NN model.
}
\label{fig:fig22}
\end{figure}

We first show that the density $\rho$ has a discontinuity across the
transition. In Fig.~\ref{fig:fig22}, we show the pdf for $\rho$ for
three different system sizes near the transition point. The pdfs have
two distinctly separated peaks that sharpen with system size. This is
a clear indication of a first order transition. 

\begin{figure}
\includegraphics[width=\columnwidth]{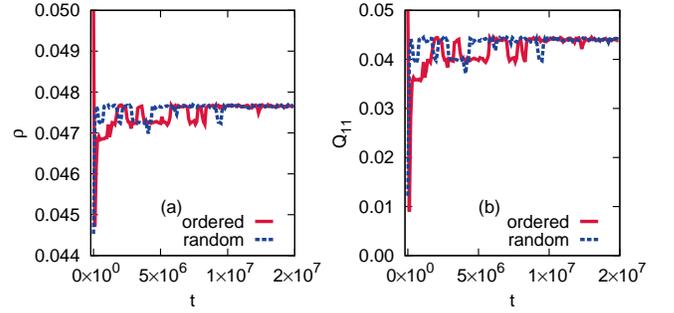}
\caption{(Color online) The variation of (a) density $\rho$ and (b) order parameter $Q_{11}$
with time $t$ for the 11-NN model when $\mu \gtrsim \mu_c$. The data
are for two different initial conditions. In the first, labelled
\textquotedblleft ordered\textquotedblright,  all particles
are in sublattice $0$. In the second, labelled \textquotedblleft random\textquotedblright, particles
are initially deposited at random.
}
\label{fig:fig23}
\end{figure}

The densities above the transition are very close to the maximum
density possible ($1/20$). Since there is a possibility that the
system may not have equilibrated, we present evidence for
equilibration. The variation of density $\rho$ and the order
parameter $Q_{11}$ with time is shown in Fig.~\ref{fig:fig23}. The
data are for two different initial conditions. In the first, the
initial configuration is one of maximal density, where all particles
are on sublattice $0$. In the second, the initial configuration is a
random one where $10^5$ deposition attempts (singe particle) are
made at random locations. We find that, though the time profiles for
the two initial conditions are different, they are statistically
identical for larger times (see Fig.~\ref{fig:fig23}(a) for $\rho$
and (b) for $Q_{11}$. Further, we also check using snapshots that the
long time behavior for both initial conditions is one every fourth
row or column are occupied. We, thus, conclude that the system is
equilibrated in our simulations.

The intermediate phase, being different from the fully packed phase,
we expect at least one more transition. But our conjecture in
Sec.~\ref{V} predicts the existence of a phase where two sublattices
are occupied. If this conjecture is true, then we expect at least two
more transitions with increasing $\mu$. Unfortunately, within
available computational time, we are unable to equilibrate the system
for larger $\mu$, and hence unable to verify the above claim.

\section{Summary and conclusions\label{VII}}

In this paper, we revisited the well-studied  two-dimensional $k$-NN hard core lattice gas
model, where the  first $k$ next nearest neighbors of a particle are   excluded from being
occupied by another particle. Using an efficient Monte Carlo algorithm, we were able to 
study numerically systems with $k$ up to $11$, earlier studies having been limited
to $k \leq 5$. Surprisingly, we found that systems with
certain $k$ ($k=4,10,11,\ldots$) undergo multiple entropy driven transitions with increasing
density. That these models may show multiple transitions have not be appreciated in 
the literature hitherto. 
For the 4-NN model, the presence of two transitions resolved an existing  puzzle as
to why the system
had a continuous transition when analogy with the Potts model predicted a first order transition. 
In this paper, we showed that the 8-fold symmetry is broken in two steps leading to two
continuous transitions -- one belonging to the Ising universality class and the other to
the two color Ashkin Teller universality class -- rather than a single first order transition. 
To rationalize this finding, we studied analytically the 4-NN model using a large $z$ expansion.
The high density phase being columnar, the expansion is in powers of $1/\sqrt{z}$ rather 
than the usual Mayer expansion in $1/z$. From the first four terms of the expansion, it
was shown that the densities of defects increases more rapidly in sublattices
where a sliding instability is present when compared to sublattices where it is absent.
This led us to conjecture that if the high density phase is columnar and the system is not
a hard square system, then the model should show multiple transitions. This conjecture 
predicts single first order transitions for $k=6,7,8,9$, and multiple transitions for 
$k=10,11$. This claim was verified numerically. 

In particular, for the 10-NN model, we showed that there are two transitions. The first
transition into a sublattice phase is continuous and is indistinguishable from the Ising
universality class. The second transition is first order. For the 11-NN model, due to 
computational limitations, we were able to numerically study only the first transition. 
However, the intermediate ordered phase was shown to be different from the phase at
full packing. Hence, at least
one more transition will definitely be present. Surprisingly, we found the intermediate 
phase to be different from what we found for the 4-NN and 10-NN models. For these latter
models, as $z$ is decreased from infinity, the system first destabilizes into a state where the
original sublattice and sublattices with sliding instability are present, consistent with
our conjecture.
Hence, we expect the 11-NN model to undergo at least three transitions with increasing 
density. 

It would be interesting to verify the above claim for 11-NN. Unfortunately, we are unable to
equilibrate the 11-NN system at densities higher than just beyond the first transition.
Seeking improvements to the algorithm in the future may help. One possible
direction would be to use flat histogram Monte Carlo algorithms like the
the Wang-Landau algorithm~\cite{wang1,wang2} or
tomographic sampling~\cite{dickman2011,dickman2014}.
A different approach would be to calculate the high density expansion for the densities of particles
in different sublattices for the 10-NN and 11-NN models. Like in the 4-NN model,
this will help to identify the preferred sublattices for generation of defects.

For the 4-NN model, by simulating systems up to $L=600$ (earlier simulations~\cite{fernandes}
having studied $L=240$), we confirm that the first transition from the low density
disordered phase to the intermediate sublattice phase is indistinguishable from the universality
class of the two-dimensional Ising model. However, sophisticated analysis based on
cluster integrals rule out Ising universality class~\cite{rotman}. This cluster analysis works 
very well for
other repulsive interaction models and is a promising tool for studying phase transitions.
It would therefore be important to understand why it fails for the 4-NN model. In particular
should  the analysis be modified in the presence of multiple transitions?

Explaining the transitions in the $k$-NN models by analytical methods is an open 
problem. Possible approaches include modified Flory approximation~\cite{fernandes2007}
and fundamental measure theory~\cite{lafuente,lafuente2003phase,schmidt} that have been
applied  earlier to HCLG models like the 2-NN model. Reproducing the two transitions 
in the  4-NN model would be a test for efficacy of these theories.  A different approach 
would be to look for
exact solutions for arbitrary $k$ on simpler lattices like the random lattice or the  Bethe 
lattice where the solution of the 1-NN model is known~\cite{runnels1nn}, or the recently
introduced random locally tree like layered lattice~\cite{rrajesh,rltl}. A precise formulation of 
what extended
hard core exclusion range means on these lattices will be the first step.
Also, to the best of our knowledge, unlike the sublattice phase, there is no rigorous proof 
for the existence of columnar phase for any model. The large $z$ expansions for the 2-NN 
model~\cite{bellemans_nigam1nn,ramola,ramola2} and the 4-NN model in this paper are only heuristic
evidence for the existence of the columnar phase. In a recent paper, the existence
of a nematic phase with orientational ordering was proved in 
Ref.~\cite{disertori}. It would
be interesting to see if such methods can be extended to prove the existence of 
columnar phase.

The algorithm used in the paper is well suited to efficiently study hard core exclusion models
on other lattices, dimensions, and for particles of different shapes. In three dimensions,
simulations of hard cubes in the continuum show a first order transition from a disordered 
phase to a
simple cubic crystal phase~\cite{smallenburg}. It would be interesting to study the lattice
version of the hard cube  model as well as the k-NN model in three dimensions, and 
obtain the phase diagram. Another interesting problem is that of rounded squares in 
two dimensions. Recent experiments on brownian squares report the existence of
hexagonal, rhombic and square phases~\cite{zhao}. Some of these features have been
reproduced in simulations of rounded squares~\cite{avendano,carmichael}. 
It is straightforward to make a lattice version
of such shapes, making it suitable to be studied by our algorithm. 
These are promising areas for future study.

\begin{acknowledgments}
We thank Deepak Dhar, Sumedha, and Anish Mallick for helpful
discussions.
The simulations were carried out on the supercomputing
machine Annapurna at The Institute of Mathematical Sciences. 
\end{acknowledgments}

%
\end{document}